\documentclass[preprint2]{aastex}

\def\sun{\hbox{$\odot$}}

\slugcomment{Submitted to the Astronomical Journal.}

\shorttitle{Larson-Tinsley Effect in the UV}
\shortauthors{Smith \& Struck}

\begin{document}


\title{The Larson-Tinsley Effect in the UV: Interacting vs.\ `Normal' Spiral Galaxies}
\shortauthors{Smith \& Struck}


\author{Beverly J. Smith}
\affil{Department of Physics and Astronomy, East Tennessee
State University, Johnson City TN  37614}
\email{smithbj@etsu.edu}

\author{Curtis Struck}
\affil{Department of Physics and Astronomy, Iowa State University, Ames IA  50011}
\email{curt@iastate.edu}



\begin{abstract}

We compare the UV-optical colors of a well-defined set of optically-selected
pre-merger interacting galaxy pairs with those of normal spirals.
The shorter wavelength
colors 
show a larger 
dispersion for the interacting galaxies than for the spirals.
This result can best be explained by higher
star formation rates on average in the interacting galaxies, combined with
higher extinctions on average.   
This is consistent with earlier studies,
that found that the star formation in interacting galaxies tends to
be more centrally concentrated than in normal spirals, perhaps due to
gas being driven into the center by the interaction.
As noted in earlier studies, there is a large variation from galaxy
to galaxy in the implied star formation rates of the
interacting galaxies, with some galaxies having
enhanced rates but others being fairly quiescent.

\end{abstract}



\keywords{galaxies: starbursts ---
galaxies: interactions--- 
galaxies: ultraviolet}

\section{Introduction}

In a ground-breaking study, Larson \& Tinsley (1978) found
that the broadband optical UBV colors of interacting galaxies have
a larger scatter than normal galaxies.  Using population
synthesis modeling, they concluded that
this scatter was caused by bursts of star formation
triggered by the interaction, superimposed on an older stellar population.
Since then, a number of studies at other wavelengths
have confirmed this basic result.
For example,
statistical studies of the H$\alpha$ equivalent widths,
H$\alpha$ luminosity per unit area,
far-infrared to blue luminosity ratios,
and mid-infrared colors
of
interacting
galaxies
imply that their mass-normalized star formation rates
are
enhanced by a factor of two on average
compared to normal spirals
\citep{kennicutt87, bushouse87, bushouse88, barton00, barton03, 
smith07}.
Statistical analyses of 
the optical spectra of large
samples of galaxies 
also support the idea that
close interactions can enhance star formation
\citep{lambas03, li08}.

With the advent of the Galaxy Evolution Explorer (GALEX)
ultraviolet telescope \citep{martin05}, a new window
on star formation in galaxies became available.
Not only is the UV a sensitive tracer of star formation,
but 
adding the UV 
to optical datasets
helps break the age$-$extinction degeneracy in population
synthesis modeling (e.g., \citealp{smith08}).
In addition,
since the UV traces 
somewhat older and lower mass stars ($\le$400 Myrs; O to early-B stars) than 
H$\alpha$ ($\le$10 Myrs; early- to mid-O stars), it provides a measure of star formation
over a longer timescale than H$\alpha$ studies.
Furthermore, since UV-bright stars are more abundant than the 
stars traced by H$\alpha$ studies, UV photometry is a better 
tool to use to study star formation rates, efficiencies,
and thresholds in regions with low gas surface density
\citep{boissier07}.
For example, GALEX observations have revealed star formation
in the outermost reaches of spiral galaxies, unseen by H$\alpha$
studies \citep{thilker05a, thilker05b, gildepaz05}.
In interacting galaxies, tidal features are sometimes quite
prominent in the UV
compared to the optical \citep{neff05, smith10}.

At the present time, the mechanisms by which
interactions trigger star formation are 
not well-understood.  
These include gas compression, shocks,
cloud collisions, threshold effects, Schmidt-type star formation laws,
stochastic processes, 
and mass transfer in and between galaxies (e.g., \citealp{keel10}).
The importance of each of these processes in star formation initiation
in interacting galaxies is not yet well-determined.
Detailed multi-wavelength analyses of individual
star forming regions within galaxies, including UV data,
in conjunction with numerical modeling of the interaction,
provides a good way to determine which processes are 
more important in an individual galaxy (e.g.,
\citealp{hancock07, hancock09, smith08, smith10, peterson09}).
To get an indication of the overall importance of each mechanism
to star formation triggering
in interacting galaxies, statistical studies of
global properties are also useful.

At present little statistical information
is available about how the global
UV$-$optical colors and the UV luminosities 
of interacting galaxies
compare to normal spirals.  
In general, interpreting the UV-optical colors of galaxies
in terms of star formation enhancements 
is more difficult than using H$\alpha$, mid-, and far-infrared 
measurements, because of an increased sensitivity to 
extinction and the fact that the UV is sensitive
to both young and intermediate-aged stars.
Furthermore, the global UV/optical colors of galaxies are strongly
affected by the distribution of dust relative to the stars,
and therefore by the morphological type of the galaxy.
To address these issues, we have conducted a 
study that involves
a large multi-wavelength database for a well-defined
sample
of nearby pre-merger interacting pairs, and a matching
dataset for a comparison sample of `normal' spirals.

\section{The Interacting Galaxy Sample and Database }

Over the last several years, we have been conducting a 
large observational
study
of a sample of more than three dozen
pre-merger interacting galaxy pairs 
(the
`Spirals, Bridges, and Tails' (SB\&T) sample;
\citealp{smith07, smith10}).
These galaxies were 
selected 
from the \citet{arp66} Atlas of Peculiar Galaxies 
to be 
relatively isolated binary systems with strong tidal distortions;
we eliminated merger remnants, close triples, and multiple systems,
unless the additional galaxies were relatively low mass.
Our systems all
have radial velocities $<$10,350 km~s$^{-1}$ and angular sizes
$\ge$3$'$.  The interacting pair NGC 4567 was added to the sample,
since it fits these criteria but is not in the Arp Atlas.
Since these galaxies are relatively simple pre-merger systems, 
they are more amendable
to the detailed matching of simulations
to observations (e.g., \citealp{struck03}) than merger remnants.
Thus this sample provides
a good testbed for investigating interaction-induced star formation.

The dataset for the SB\&T sample includes mid-infrared imaging
with NASA's Spitzer space telescope \citep{smith05,
smith07, hancock07}
as well as ultraviolet imaging
with GALEX 
\citep{hancock07, smith10}.
In addition, 
broadband {\it ugriz} optical images are available for most
of the sample galaxies from the Sloan Digitized Sky Survey (SDSS; 
\citealp{abazajian03}).
We have already produced an Atlas of the Spitzer infrared images
\citep{smith07}
and a second Atlas of the GALEX and SDSS images
\citep{smith10}.
In an earlier study \citep{smith07},
we compared the Spitzer broadband mid-infrared
colors of this sample with those of a control sample
of `normal' spirals selected
from the 
\citet{kennicutt03}
SINGS sample by eliminating strongly interacting galaxies.
Multi-wavelength studies of individual galaxies in the sample
have been presented by \citet{hancock07, hancock09}, \citet{smith05, smith08},
and \citet{peterson09}.

\section{A Large `Control' Sample of Spirals}

The next step is to 
test whether the UV emission from our sample galaxies
is also enhanced by the interactions, 
and if there is a statistically-significant
difference in the UV-optical colors of our galaxies 
compared to spirals.  
Unfortunately, however,
only a subset of the `control' sample of `normal'
spirals we used
in our 
earlier Spitzer study \citep{smith07} 
has reliable optical magnitudes available
\citep{munoz09},
and the final sample size is too small for a valid comparison.
No other appropriate control sample with both UV and optical data is 
currently available in the literature.

To address this lack, we have constructed a new control sample.
To ensure the availability of high quality UV data ($\ge$ 1 orbit observing
time with GALEX),
we started with the 1034 galaxies
in the `GALEX Ultraviolet Atlas of Nearby Galaxies'
\citep{gildepaz07}.
This contains most of the 
large angular size
galaxies in the local Universe 
that have been 
observed by GALEX to date.
We selected the subset of these galaxies 
classified as Sa $-$ Sd spirals in the \citet{gildepaz07}
tabulation, and searched the NASA Extragalactic
Database (NED\footnote{The NASA Extragalactic Database;
http://nedwww.ipac.caltech.edu}) for companions.  We 
eliminated galaxies with companions that
have velocity differences $<$ 1000 km~s$^{-1}$, separation on
the 
sky $<$ 10 $\times$ (diameter of target + diameter of companion),
and optical magnitude(companion) $<$ 1.5 + magnitude(target galaxy).
We then limited the resultant sample to galaxies with distances between
6.2 $-$ 143 Mpc, the range of distances in the SB\&T sample.
We then omitted
galaxies with angular sizes larger than the SDSS field of view of $\sim$8$'$
and galaxies split between two SDSS or GALEX images.
We inspected the images for 
previously-unidentified morphological 
peculiarities
and
companions, as well as    
artifacts in the images.  We eliminated such galaxies, leaving 
a final sample of 
121 galaxies. 
Of these galaxies, only five are listed as possible Seyferts in
NED, compared to nine out of the 84 main galaxian disks in the 42
SB\&T pairs.

As noted above, we selected our spiral sample to have the
same range of distances as the SB\&T sample.   This is
shown in Figure 1, where we provide histograms of
the distances to the galaxies in the two samples.  
On average, the spirals are slightly more distant, with
a median distance of 68 Mpc, vs.\ 37 Mpc for the SB\&T galaxies. 
A K-S test give a probability
that the two sets of distances are drawn from the same
parent population of 3.3\%, thus the difference between
these two samples is
marginally significant.

This sample and the SB\&T sample both contain mostly high latitude
galaxies. 
The distributions of Galactic latitudes {\it b} for the two
samples are shown in Figure 2.  Almost all the galaxies
have $|${\it b}$|$ $>$ 25$^{\circ}$, with the median $|${\it b}$|$ 
being 54$^{\circ}$ for the SB\&T sample and 
48$^{\circ}$ for the comparison spirals.
A K-S test gives a probability of 1.1\% that the two distributions
come from the same parent sample, thus there is a significant difference
between the two samples.   However, this is unlikely to have a large
effect on our results, since at these latitudes the corrections
for Galactic extinction are small.

\section{The Large-Scale Environment }

In addition to the presence of nearby neighbors, 
another factor that affects galaxian
star formation rates 
is the large scale environment, i.e., whether
the galaxy is in a group, cluster, or the field 
(e.g., \citealp{gomez03}).  
Therefore,
in constructing a comparison
sample, it is important to also compare the large scale environment
of the galaxies 
(e.g., \citealp{alonso06, barton07}).
To this end, for our sample `normal'
galaxies, we have used NED to determine
the number of neighboring galaxies.
We used a search radius of 2 Mpc, a velocity difference
of $\le$ 1000 km~s$^{-1}$, and an absolute magnitude limit
for the neighbor of M$_{\rm V}$ = $-$18.5 (approximately equal
to that of the Large Magellanic Cloud).
This search was constrained by the maximum
search radius in NED of 300 arcminutes,
thus is deficient for very nearby galaxies ($<$23 Mpc), however,
this only affects a small number of galaxies in the two samples 
(see Figure 1).

Histograms of the number of neighbors 
that fit these criteria
are given in Figure 3
for the two samples of galaxies.
This figure shows four outliers in the SB\&T sample, in very
dense environments.  In order from right to left, these are
Arp 105 (in the Abell 1185 cluster), Arp 120 (in the
Virgo cluster), Arp 65 (in the WBL 009 cluster), and NGC 4567 (in Virgo).
Other than these outliers, the two distributions are similar.
A KS test gives a 44\% chance that the
two distributions arise from the same parent population, thus 
environmental
differences between the two samples are statistically insignificant.

\section{Morphological Types }

To better interpret any possible differences in UV-optical colors between
these two samples, 
we first
compare the morphological types of
the individual galaxies in the samples.   
To this end, we used NED
to extract morphological 
types for the individual galaxies
in the SB\&T pairs, as well as for the spiral sample.   
These types are plotted as histograms in the top and middle
panels of Figure 4.
For these histograms, we ignored any notes in NED
on morphological peculiarities 
(i.e., Sab Pec), and just used the basic Hubble type (i.e., Sab).
The fact that many of these galaxies are peculiar, combined with 
the fact that 
the types in NED are inhomogeneous and 
subjective,  means that there are significant uncertainties in the types
plotted in Figure 4.
However, to first order these histograms suggest that the
two sets of galaxies are different morphologically,
in addition to 
the presence of tidal features in the SB\&T galaxies.
For example,
the SB\&T sample contains quite a few galaxies classified as
S0 and Im/Irr/Sm, as well as a few ellipticals.
In contrast, the spiral sample was intentionally selected to
avoid these types.  
Furthermore, the distribution of types for the SB\&T galaxies
peaks
at Sb, while that for the spiral sample peaks at Sc/Scd.
This suggests a possible difference in the bulge/disk ratios
of the two samples.

The two samples appear to have similar fractions of barred spirals.
About 30\% of the spirals in the interacting sample are 
classified as SB in NED,
compared to 33\% in the comparison spiral sample.  
For the interacting sample, 14\% of the spirals are listed as SAB in
NED, compared to
19\% of the comparison sample galaxies.

Since a strong interaction can change the appearance
of a galaxy, the morphological types in NED do not 
necessarily reflect the original pre-interaction structures of the
galaxies.
Unfortunately, after the fact it is often difficult to tell original
morphological types.   
It is unclear whether the skewing of the spiral subset of the SB\&T 
sample to earlier Hubble types is due to a difference in progenitors
caused by selection effects,
to changes produced by the interaction itself, or
to biases in the typing of the interacting galaxies.

To partially address the issue of morphological bias in the samples, 
in the following analyses 
we only include the SB\&T
galaxies with current morphological types in NED between S0/a and Sdm. 
Since the SB\&T spirals are skewed to 
earlier Hubble
types compared to the comparison spirals, 
in addition to comparing to the full comparison sample,
we have repeated the
analysis using Hubble-type-matched subsets of the comparison sample. 
These subsets were constructed by randomly selecting galaxies 
from
the comparison sample to fit a fixed distribution of 
Hubble types which
approximately matches the Hubble type distribution of the
spirals in the SB\&T sample.  This fixed distribution is plotted
in the third panel of Figure 4.  These subsets have
61 galaxies each, thus contain approximately half of the 
original sample of comparison spirals.   We randomly selected 100 such
subsets, and ran the statistical tests with each subset, to
investigate how much random chance affects the final statistics. 
These results are described below.

\section{UV/Optical Magnitudes of Spiral Sample }

For the galaxies in our spiral sample,
we 
used the GALEX and SDSS images to extract total
FUV, NUV, and {\it ugriz} magnitudes, 
using the same
method as we used for the SB\&T galaxies
\citep{smith07, smith10},
including an additional term
in the uncertainties to account for sky variations. 
Of the 121 spiral
galaxies in our sample, 12 do not have FUV images available,
thus FUV magnitudes are not available for these systems.
We corrected the observed magnitudes for Galactic extinction
in the same way
as for the SB\&T galaxies \citep{smith10}.
The corrected magnitudes are given in Table 1.

\section{Luminosity Comparisons }

The next question is whether the luminosities and stellar masses
of the two samples differ.
In Figures 5 and 6 we plot the NUV and g luminosity distributions of
the SB\&T spiral sample and the full comparison sample, with 
the luminosities
being calculated as $\nu$L$_{\nu}$.   
For the SB\&T sample, we plot the luminosities of the main disks
separately from those of the tails and bridges.  We 
also
plot separately the luminosities of the candidate tidal dwarf galaxies (TDGs),
as identified in \citet{smith10}.
Tidal dwarf galaxies are concentrations of young stars and gas in tidal
features that may or may not evolve into independent dwarf galaxies.

The median fraction of the total galaxian
luminosity in the tidal features plus that
in the TDGs
is $\sim$15\% in the UV bands, and $\sim$10\% in the optical.
However, this fraction varies widely from galaxy to galaxy,
ranging from $<$1\% $-$ 65\% in the UV bands, and $<$1\% $-$ 40\% in
the optical.
For comparison, 
for a different subset of Arp Atlas galaxies,
\citet{schombert90} found an average of
$\sim$25\% of the total visible starlight was contained
in tidal features.
For the SB\&T galaxies, in the Spitzer 3.6 $\mu$m $-$ 8 $\mu$m
bands we found $\le$10\% of
the light was 
coming from the tidal features on average \citep{smith07}.

The NUV luminosities for the disks of the comparison spirals are very slightly
larger than for the SB\&T spirals on average, with
median luminosities
of
10$^{43.0}$ erg~sec$^{-1}$ and
10$^{42.8}$ erg~sec$^{-1}$, respectively.
If the light from the tidal features were combined
with that of the SB\&T disks, this would
partially account for the
difference.
A
K-S test give a probability that these two sets of 
luminosities
came from the same
sample of 33\%, thus we cannot rule out the hypothesis
that the two sets of data come from the same parent
population.
The median g luminosity for both the comparison spirals
and the SB\&T spirals are 10$^{43.5}$ erg~sec$^{-1}$, with
a 
K-S test giving a probability of the two parent populations
being the same of 28\%.   Thus the g luminosities of the two
samples are also
not significantly different.   The small difference
in the distribution of distances for the two samples therefore does
not produce a strong difference in luminosities.
We conclude that there is likely not a large
difference in the mass distribution for the two samples.

\section{Color-Color Plots}

The main result of the \citet{larson78}
study
was that interacting galaxies show a larger scatter in the UBV color-color
diagram than normal galaxies.   
To search for this effect in our dataset and to extend to other wavelengths,
in Figures 7 $-$ 11 we provide
various color-color plots for the full comparison sample of 
spirals (right panel),
compared
to the SB\&T disks, excluding the E/S0/Sm/Im/Irr galaxies 
(left panel).  
Following \citet{larson78}, on these plots we include curves
showing the running averages of the x-value for the comparison spirals
(solid curves) and SB\&T spirals (dotted curves)
as a function of the y-value,
calculated in bins of 0.1 magnitudes,
except for NUV $-$ g, where we used 0.2 magnitude bins.
These curves help guide the eye to see differences between the samples.

We calculated the rms deviation of each dataset relative to these mean 
lines.   These values are given in Table 2.
For the SB\&T galaxies, we calculated the rms deviation relative
to the mean curve for the SB\&T sample itself (fourth column
in Table 2), as well as the rms deviation relative
to the mean curve for the full comparison spiral sample
(fifth column).
For the 
FUV $-$ NUV vs.\ g $-$ r, NUV $-$ g vs.\ g $-$ r,
and u $-$ g vs.\ g $-$ r plots (Figures 7 $-$ 9), the 
rms deviations of the SB\&T spirals are indeed larger than for 
the comparison spirals, especially when calculated compared to the
mean line for the comparison spirals.
In contrast,
for the two longer wavelength
color-color plots (Table 2 and Figures 10 and 11), 
there is little difference in
the scatter, as expected for redder colors dominated mainly by the
light from older stars.

We repeated this analysis with the Hubble-type-matched subsamples,
and found similar results.   In Table 2, we provide the mean rms deviations
for our 100 randomly-selected subsets.    On average,
for the shorter wavelength colors,
these subsets also show smaller scatter than the 
SB\&T galaxies.

Inspection of these plots shows a few additional
differences between the two samples.
First, there is a subset of the SB\&T spirals with significantly
bluer NUV $-$ g
colors for their g $-$ r values (Figure 8).
Second, in Figure 9, the u $-$ g vs.\ g $-$ r plot,
the datapoints for the interacting sample tend to lie to the right
of the mean line for the comparison spirals (i.e., to 
{\it redder} u $-$ g values for a given g $-$ r).
In g $-$ r vs.\ r $-$ i (Figure 10), a similar effect is 
seen: there is 
a slight tendency of the interacting sample towards redder g $-$ r values compared to r $-$ i.
However, as noted above, the scatter relative to the mean spiral line
for g $-$ r vs.\ r $-$ i
is similar in both samples, so any systematic offset is small.
In the other two color-color plots, FUV $-$ NUV vs.\ g $-$ r (Figure 7) 
and r $-$ i vs.\ i $-$ z
(Figure 11), no systematic offset between the two samples 
is evident within the uncertainties.
Inspection of similar
color-color plots for the Hubble-type selected subsets
shows similar trends.

\section{Color-Morphology Relations }

As noted earlier, there is a difference in the morphological types
of the galaxies in the two spiral samples, with more Sb galaxies
in the SB\&T galaxies, and more Sc galaxies in the 
full comparison spiral set.
To investigate further the relation between
morphology and color,
for both the SB\&T galaxies and the full
comparison sample,
in Figures 12 $-$ 17
we compare the morphological types with
FUV $-$ NUV, NUV $-$ g, u $-$ g, g $-$ r,
r $-$ i, and i $-$ z, respectively.
Clear trends are seen in these plots, in that earlier Hubble types
are redder, as expected (e.g., \citealp{deV77, gildepaz07}).   

For each spiral Hubble type and each color
for both sets of galaxies, 
we have
calculated the mean color.
These mean value curves are plotted on Figures 12 $-$ 17, with
the solid curve being the mean values for the comparison sample,
and the dotted curve the SB\&T sample.
In Table 3, we give 
the rms deviations of the datapoints relative to these curves.
For the three shorter wavelength colors,
these residuals are larger for the 
SB\&T sample, reflecting either a larger range in 
star formation rates
and/or
extinctions for a given Hubble type for the interacting
sample, and/or larger inaccuracies in Hubble typing.

At the shorter wavelengths,
for some of the Hubble types 
there is a tendency for
the mean colors of the interacting galaxies to be slightly
redder than for the comparison spirals.  However, this
effect is slight.   The main difference between the two samples
is the somewhat larger scatter for the interacting galaxies.

We repeated this analysis for the Hubble-type-matched subsets
of the comparison sample, and find similar results.  
The mean values for 100 type-match subsets
are also given in Table 3.

\section{Comparison with Previous Studies of Optical Colors}

We find a significant difference in the optical
colors of our spiral and interacting
samples, with the interacting galaxies showing more dispersion
in the colors.
Our study thus
confirms the basic results of \citet{larson78},
who saw greater scatter in the UBV optical colors for their sample
of Arp galaxies than for galaxies from the Hubble Atlas
of Galaxies \citep{sandage61}, using photometry from the Second Reference
Catalogue of Bright Galaxies (\citealp{deV76}; RC2).
However,
there are some differences between that
study and the current
study.  In their UBV color-color plot, \citet{larson78} saw more scatter to
the blue in U $-$ B for a given B $-$ V for the interacting
galaxies compared to their control sample.  In contrast,
in our u $-$ g vs.\ g $-$ r plot, 
our interacting spiral disks lie to {\it redder} u $-$ g colors
than the comparison spirals
for a given g $-$ r color.
This could be due in part to differences in the filters as well as 
differences in sample selection.
We intentionally selected spirals for our
control sample, while \citet{larson78} used the full
RC2, which includes ellipticals as well.   Second, they used
the full Arp Atlas as their initial sample of interacting
galaxies, while we selected a subset of pre-merging
binary pairs.   Our interacting
sample thus omits merger remnants, while theirs
does not.
We also eliminated galaxies typed as irregulars from our `spiral
only' interacting sample, while \citet{larson78} did not.

There have been a few previous follow-up studies to the
\citet{larson78} paper.
\citet{schombert90} imaged a set of 25 Arp Atlas galaxies
in the BVri optical filters, and found a bigger scatter in
the BVr color-color diagram for the disks of their Arp galaxies 
compared to a normal galaxy sample.   However, they did not
note a pronounced shift to either the blue or the red in these colors.
\citet{bergvall03} compared optical 
colors of a magnitude-limited
sample of 59 interacting galaxy pairs with a similar sample of galaxies
without massive companions.   
In both their interacting and non-interacting samples, $\sim$30$\%$
of the galaxies are classified as ellipticals, in contrast to our samples,
for which ellipticals were eliminated.  As with the other studies discussed
above,
\citet{bergvall03} also find a slightly large dispersion
in the UBV colors of the interacting sample compared to the non-interacting
galaxies.   They 
also found a slight offset in the locations of the two samples on the UBV
color-color diagram,
with the distribution of colors for the interacting galaxies shifted
to bluer B $-$ V and redder U $-$ B relative to the locus
of the non-interacting systems.
This is similar to our u $-$ g vs.\ g $-$ r plot (Figure 9).
In another study, \citet{alonso06} compared optical SDSS u $-$ r 
and g $-$ r
colors of close pairs and non-pair galaxies in the same
large-scale environment, and found excesses of both
red and blue galaxies in the pair sample.

The consistent result from all of these studies including our
own is that there is somewhat
larger scatter in the optical colors for the interacting galaxies.
This is not a large effect, likely because 
the relationship between star formation and optical
colors is a complex function of age and extinction.
This point is discussed below.
In addition,
in some sets of colors and some datasets there is a 
systematic shift in color between
the two samples.  Whether or not such a shift is detectable,
and in which direction, appears to
depend to some extent upon sample selection, filters, and/or sample size.


\section{Comparison with Stellar Population Synthesis Models}

In contrast to earlier studies using H$\alpha$, 
mid-infrared, or far-infrared
observations,
interpreting UV-optical colors in simple terms of average
star formation enhancements
is more difficult, 
both because older stars contribute more to powering
the UV/optical fluxes from galaxies, and 
because of the larger effect of extinction
in the UV and short wavelength optical. 
Because of these factors,
only very weak correlations are seen between the 
global UV/optical
colors of our SB\&T galaxies and their Spitzer [3.6] $-$ [8.0]
and [3.6] $-$ [24] colors \citep{smith10}.
As discussed earlier, the biggest difference between the UV/optical
colors of our two samples is not a systematic shift in these 
colors, but a larger scatter in the colors of the interacting sample.

To explain these differences, we made some simple comparisons with
population
synthesis models.  
In Figures 18 $-$ 22, we again display the
various color-color plots for the two spiral samples.
On these plots, 
we have superimposed evolutionary
tracks from version 5.1 of the
Starburst99 (SB99)
population synthesis code \citep{leitherer99}.
Following \citet{larson78},
in these models
we have combined an underlying
older
stellar population with a burst of star formation.
We calculated models with a range of `burst strengths' 
f$_z$(young)/f$_z$(old) = 0.01 (green dotted curves), 0.1 (blue short
dashed curves), and
0.2 (red long dashed curves), 
where
f$_z$(young)/f$_z$(old) 
is the flux density in the z band due to the young stellar population,
compared to that from the older stars.
In Figures 18 $-$ 22, the plotted curves
represent models with increasing
burst age and constant burst strength.
For the underlying
older
stellar population for these models, we have used moderately red
colors that lie 
along the mean color curves for the comparison spiral sample,
at 
FUV $-$ NUV = 0.75, 
NUV $-$ g = 2.6, 
u $-$ g = 1.3, 
g $-$ r = 0.6, 
r $-$ i = 0.27, and
i $-$ z = 0.27.
Thus we are assuming that the pre-interaction galaxies are typical 
star-forming spiral
galaxies which have undergone an additional burst of star formation due 
to the interaction.
We note that if the assumed underlying population is
different than
these nominal values, the general shape of the curves in Figures
18 $-$ 22
will remain the same,
but they will be shifted to the new endpoint and the curves
will be compressed, for a blueward shift of the underlying
population,
or
stretched out for a redward shift.

For these models, we have
assumed an instantaneous burst,
solar metallicity, 
\citet{kroupa02}
initial mass functions (IMF), and an initial mass range of 0.1 $-$ 100
M$_{\sun}$.
This version of the SB99 code 
includes the Padova asymptotic giant-branch
stellar models \citep{vaz05}.
We calculated model colors for a series of
ages starting at 
1 Myrs, increasing by step sizes of 1 Myr to 20 Myrs,
then by 5 Myr steps to 50 Myrs, then 10 Myr steps to 100 Myrs.
For all models, we included the nebular continuum from SB99.
As in \citet{smith08, smith10}, 
we also added the H$\alpha$ line 
to the model flux for the r band filter.
For very young bursts (1 $-$ 5 Myrs),
the r magnitude will decrease
by 1.1 $-$ 0.25 magnitudes due to 
H$\alpha$, causing a sudden `bend' in the model curves
to redder g $-$ r
at young ages.
This bend is visible in the FUV $-$ NUV vs.\ g $-$ r plot (Figure 18),
at young burst ages.
For the calculations of the H$\alpha$
contributions, we used the redshift of Arp 285, 
which is typical
of the SB\&T sample.  For the
highest redshift galaxies in our sample, this will be an over-estimate, as
H$\alpha$ will be redshifted to a less sensitive
portion of the r band response function.
Note that our models do not include the [O~III] $\lambda$5007 or
H$\beta$ line, which can contribute substantially to the g filter
for low metallicity young galaxies (e.g., \citealp{kruger95, west09}),
thus partially counteracting the reddening trend in g $-$ r.

In Figures 18 $-$ 22, we have also included arrows showing some possible
reddening vectors due to dust.   The amount of reddening in each
band depends upon the type of dust and the geometry of the system,
and can differ quite a bit depending upon the assumptions made.
To illustrate the range in possible reddening models,
on these plots we have included three example reddening vectors
from the multiple-scattering
radiative transfer
calculations of \citet{witt00}.
In all three models plotted, the length of the vector
plotted corresponds to ${\tau}_{\rm V}$ = 1.  The solid arrow corresponds
to a model with a smooth distribution of Small Magellanic Cloud-like dust 
evenly mixed amongst the stars.  The dashed arrow corresponds to clumpy
Milky Way-like dust mixed with the stars.  The dotted arrow 
is produced by clumpy SMC dust distributed in a shell outside of the stars.
See \citet{witt00} for more information about these models.

We recognize that the simple population synthesis models plotted in Figures 18 $-$ 22
do not accurately represent the true
star formation histories of these galaxies. First,
the older stellar population
likely varies from galaxy to galaxy, which will
shift the location of the model curves shown in Figure 18 $-$ 22.
Second, the young stellar
population may not be best represented by an 
single instantaneous burst.
Also, the geometry of the system is likely more complex than in these
simple reddening models,
and the extinction may be different
for the 
young and old stellar populations.
However, these models are still valuable for illustrating some general trends.
Our goal is not to provide accurate star formation histories for
individual galaxies in the two samples, but simply to look for 
trends that will help in interpreting the differences between
the two samples.

In all of these color-color plots, some
degeneracy between age and extinction is visible, as seen by 
a comparison of the burst models
and the extinction vectors.  
When bursts of different strengths are added to 
the older stellar population,
the models shift in characteristic ways in the color-color plots.
For relatively young bursts,
the FUV and NUV bands
are dominated by light from the burst.
In contrast, the i and z optical bands are predominantly 
light from the older stars.  The u, g, and r bands have 
important contributions
from both, with the proportions varying with burst strength and age.
In the FUV $-$ NUV vs.\ g $-$ r diagram (Figure 18), 
increasing burst strength causes a shift in g $-$ r
towards the blue.
In this plot, the age and burst strength vectors
point in different directions.
In the NUV $-$ g vs.\ g $-$ r plot (Figure 19),
increasing
the burst strength also causes a shift
at an angle to the direction of increasing burst age.
In contrast, in the two longer wavelength
optical plots (Figures 21 $-$ 22), increasing the 
burst strength shifts both colors
bluewards, and the direction of increasing burst strength is 
approximately parallel
to the age vector, and 
roughly
anti-parallel with the extinction vector.
Thus there is more degeneracy in these two plots.

In the shorter
wavelength plots, the main difference
between 
the two samples is the less concentrated distribution of points in the 
interacting
galaxy sample, without a strong difference in the medians
of the colors for the two samples.  
Stronger bursts alone would tend to shift the medians to
bluer colors,
while larger extinctions alone would shift the medians to
the red.
The lack of a pronounced systematic shift in colors between
the two samples 
suggests that, on average, 
the interacting sample has both stronger `bursts' in some galaxies
(i.e., higher current rates of star formation),
and larger extinctions in others.  These two effects together
will tend
to spread out the distribution of colors in the observed manner 
(see Figure 18).

As noted earlier,
in the NUV $-$ g vs.\ g $-$ r plot (Figure 8), there is a slight shift
to bluer NUV $-$ g for a given g $-$ r for the interacting galaxies,
in addition to larger scatter.   This could be accounted for by
the combined effect of some galaxies having stronger bursts, and other galaxies
having stronger extinctions (see Figure 19).
In the u $-$ g vs.\ g $-$ r plot (Figure 9), there is a shift to 
bluer g $-$ r
values for a given u $-$ g value (or equivalently, a reddening of u $-$ g
for a given g $-$ r).  A shift in that direction can be produced by
stronger bursts (Figure 20), on average, for the interacting galaxies.
However,
to explain the reddest colors in this plot requires some of the
interacting galaxies to have more extinction,
on average, compared to the comparison sample.

Thus we conclude that the interacting galaxies, on average, have
{\it both} stronger bursts compared to the comparison
sample, {\it and} higher average extinctions.   However, there are
strong variations across the sample in both burst strength
and in extinction.  This is consistent with earlier studies based on
H$\alpha$ and far-infrared observations
\citep{kennicutt87, bushouse87, bushouse88},
which showed that the 
enhancement in
the star formation rate of interacting galaxies is an average effect;
some interacting galaxies show no enhancement at all.  

Our conclusion that the 
interacting galaxies tend to have stronger extinction on average
is consistent with earlier studies which 
found that the star formation in interacting galaxies tends to
be more centrally
concentrated than in normal spirals \citep{bushouse87, smith07}.
Numerical models suggest that interactions can drive gas into the central
regions of galaxies, triggering central starbursts.
This central concentration
may lead to larger extinction on average in the interacting sample.
In support of this idea, \citet{bushouse87} and \citet{bushouse88} found
higher 
L$_{FIR}$/L$_{H{\alpha}}$ ratios in their interacting sample compared
to spirals, also suggesting more obscured star formation on average.

In addition to age, extinction, and burst strength variations, other 
factors that can cause scatter in the observed color-color diagrams are
galaxy-to-galaxy
variations in the IMF, the metallicity,
the properties of the dust, particularly 
the dust attenuation law, and 
the 
spatial distribution of the dust relative to the stars, including
the clumpiness of the dust.
The relative importance of these factors to the observed
colors of galaxies is discussed in detail by \citet{conroy10a, conroy10b}.
A clumpier dust distribution will tend to make the observed
colors bluer, while
a steeper IMF will redden the colors.
At the present time, however, there is no strong evidence of 
significant differences between the two samples in any of these parameters.

\section{Summary }

We have compared the UV/optical colors of a well-defined sample
of pre-merger interacting galaxy pairs with those of `normal spirals'.
As previously found for optical colors alone by \citet{larson78},
the interacting galaxies show a larger dispersion in their colors
compared to the normal spirals.  
This effect is especially pronounced in the shorter wavelength
UV/optical colors, 
due to the increased sensitivity to both young stars and 
dust.  This increased scatter is best explained
by a combination of both moderately-enhanced star formation 
in the interacting galaxies, along with larger average extinction.
In interacting systems, gas may be driven into the central
regions of the galaxies, triggering star formation while
at the same time increasing the average column density of dust.
While massive young stars are blue, they tend to be reddened
by heavy layers of dust, thus interacting galaxies are not dramatically
bluer on average than non-interacting systems.
However, the dust screens are known
to be irregular or fractal, so some very blue light can escape.
These combined effects tend to increase the scatter in the observed
UV/optical colors of interacting systems.

Our multi-wavelength
database for both the SB\&T sample
and the 
`normal' spiral galaxies
should 
prove useful for
future comparisons with other galaxy samples
such as cluster spirals and spirals in compact groups, to investigate
star formation enhancement in these environments.   
Both samples should also
be useful for comparison with high
redshift galaxies.



\acknowledgments

We are very grateful to the GALEX and SDSS teams for 
making this research possible.
We thank Mark Hancock for his earlier work on population synthesis modeling.
We also thank Amanda Moffett for some preliminary work related
to this project.
We thank the anonymous referee for suggestions that greatly improved
this paper.
This research was supported by GALEX grant GALEXGI04-0000-0026,
NASA LTSA grant NAG5-13079, NASA Spitzer
grants RSA 1353814 and RSA 1379558, and NASA Chandra grant AR9-0010A.
This research has made use of the NASA/IPAC Extragalactic Database (NED) 
and the NASA/ IPAC Infrared Science Archive, 
which are operated by the Jet Propulsion Laboratory, 
California Institute of Technology, under contract with the National 
Aeronautics and Space Administration.

\clearpage



\clearpage

\begin{figure}
\plotone{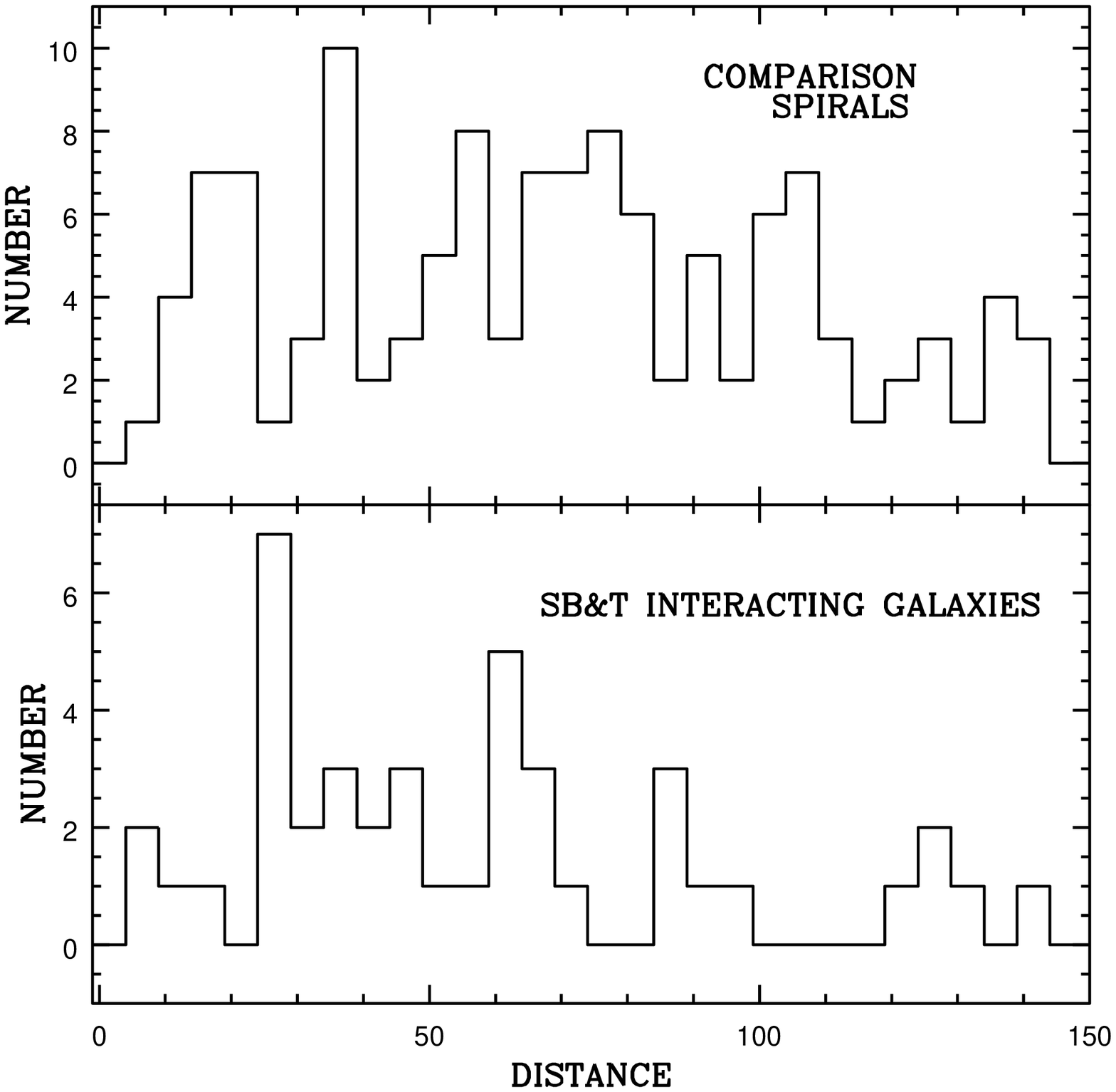}
\caption{
  \small 
The distribution of distances for the galaxies in the two samples.
}
\end{figure}

\begin{figure}
\plotone{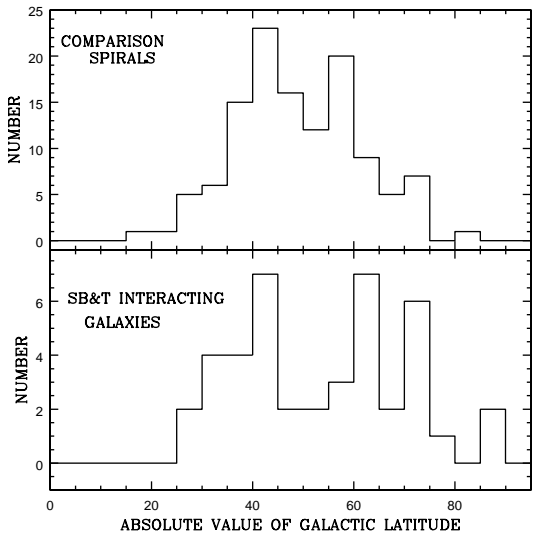}
\caption{
  \small 
The distribution of Galactic latitudes for the two samples.
}
\end{figure}

\begin{figure}
\plotone{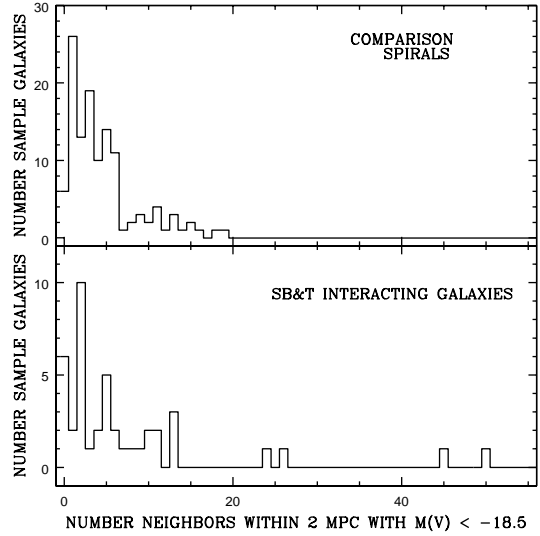}
\caption{
  \small 
The number of neighbors per galaxy in each sample, when a neighbor
is defined as being within 2 Mpc and $\delta$V $\le$ 1000 km~s$^{-1}$,
with an absolute V magnitude brighter than $-$18.5.
}
\end{figure}

\begin{figure}
\plotone{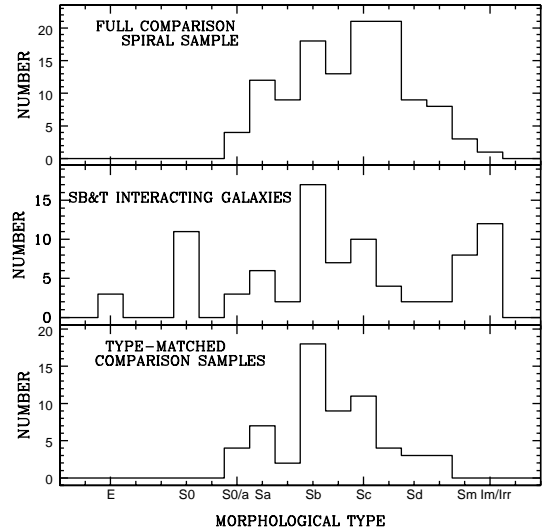}
\caption{
  \small 
The distribution of morphological types of the two samples,
as well as that used for the type-matched subsets of the comparison
sample.
See text for more information.
}
\end{figure}

\begin{figure}
\plotone{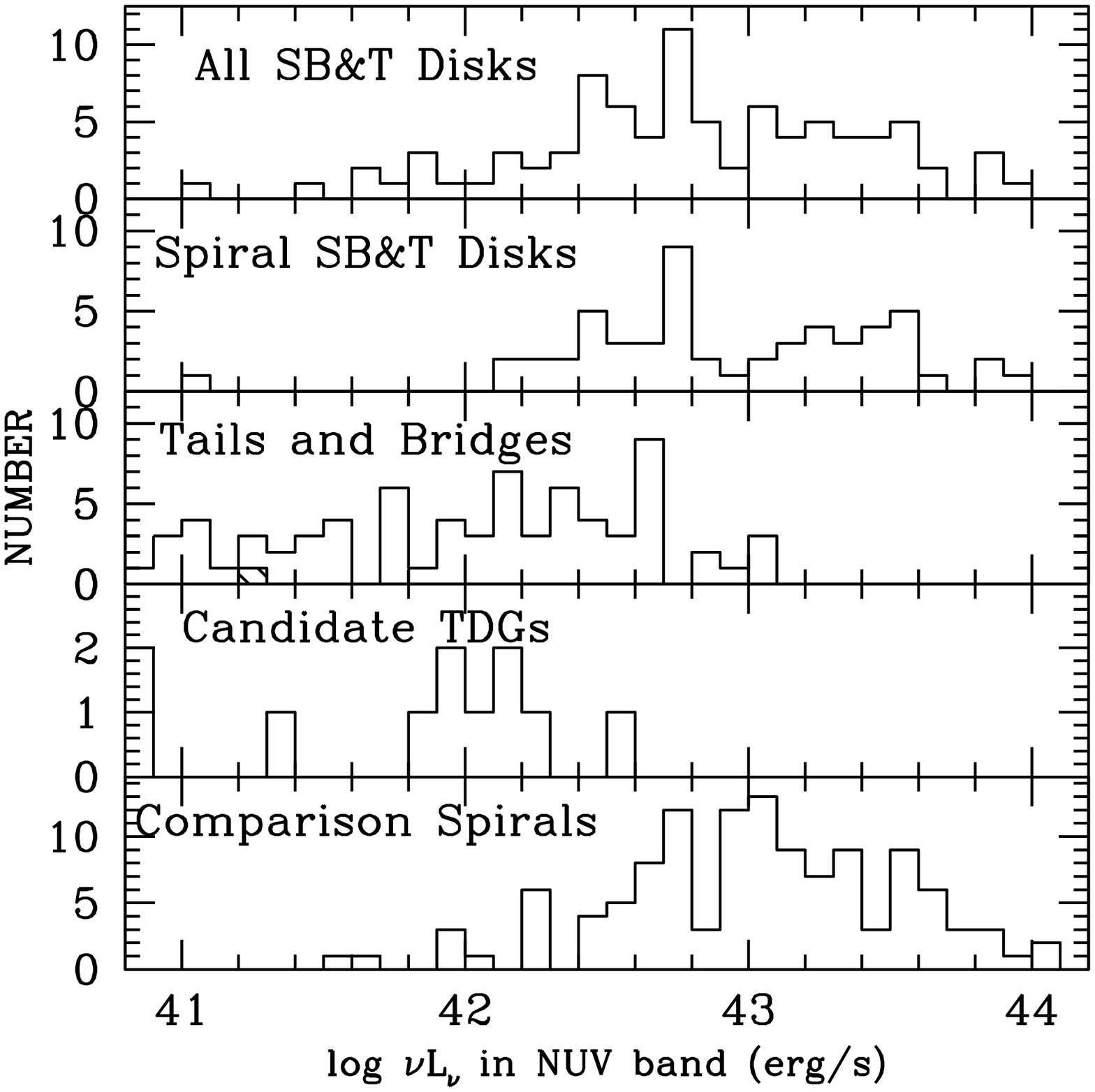}
\caption{
  \small 
A comparison of 
the distributions of NUV luminosities for the full comparison
sample (bottom panel), the full SB\&T sample (top panel),
the SB\&T spiral-only sample (second panel), the tails
and bridges (third panel), and the candidate tidal dwarf galaxies
(fourth panel).
The hatched regions are upper limits.
}
\end{figure}

\begin{figure}
\plotone{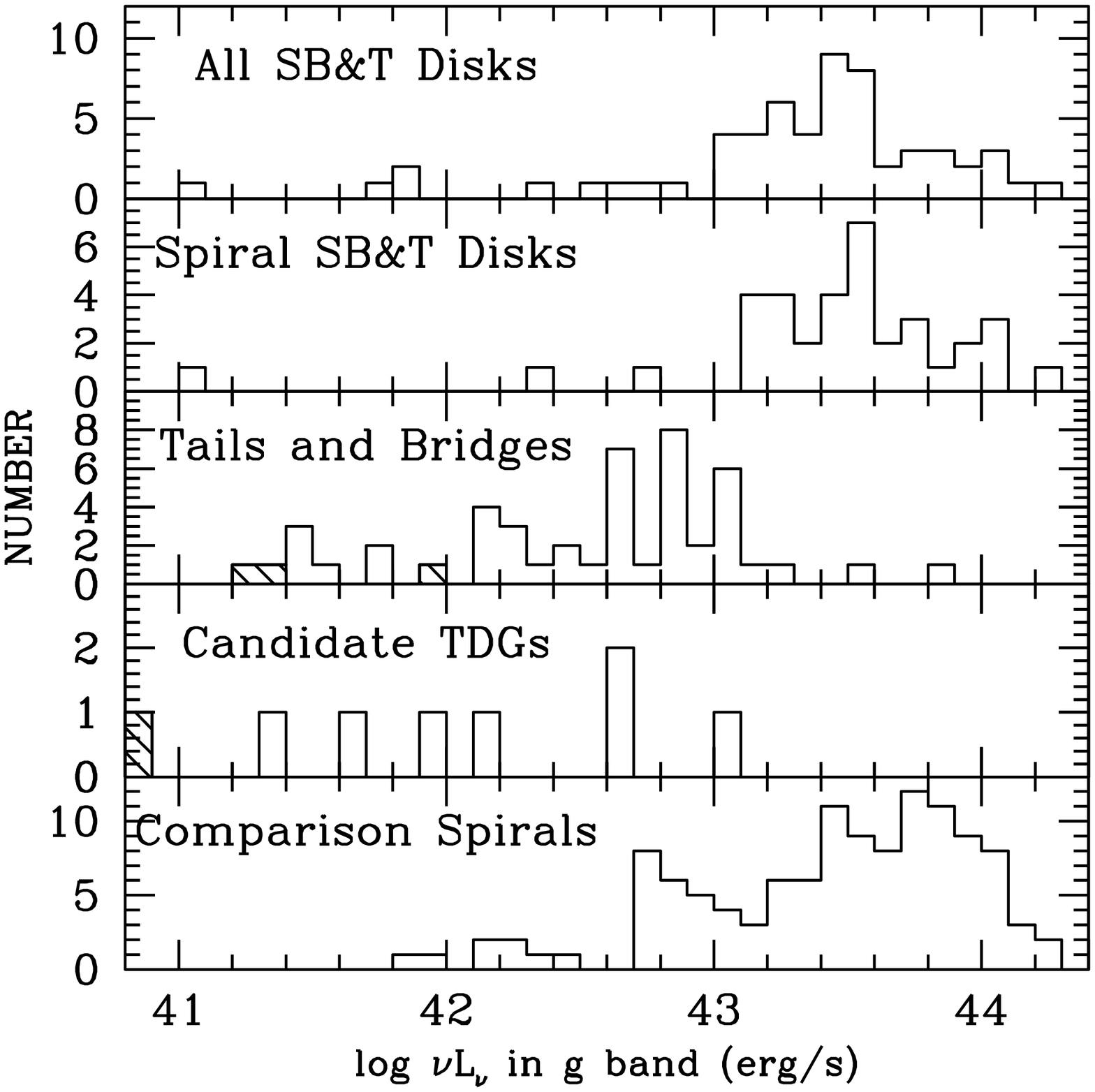}
\caption{
  \small 
A comparison of 
the distributions of g luminosities for the full comparison
sample (bottom panel), the full SB\&T sample (top panel),
the SB\&T spiral-only sample (second panel), the tails
and bridges (third panel), and the candidate tidal dwarf galaxies
(fourth panel).
The hatched regions are upper limits.
}
\end{figure}

\clearpage

\begin{figure}
\plotone{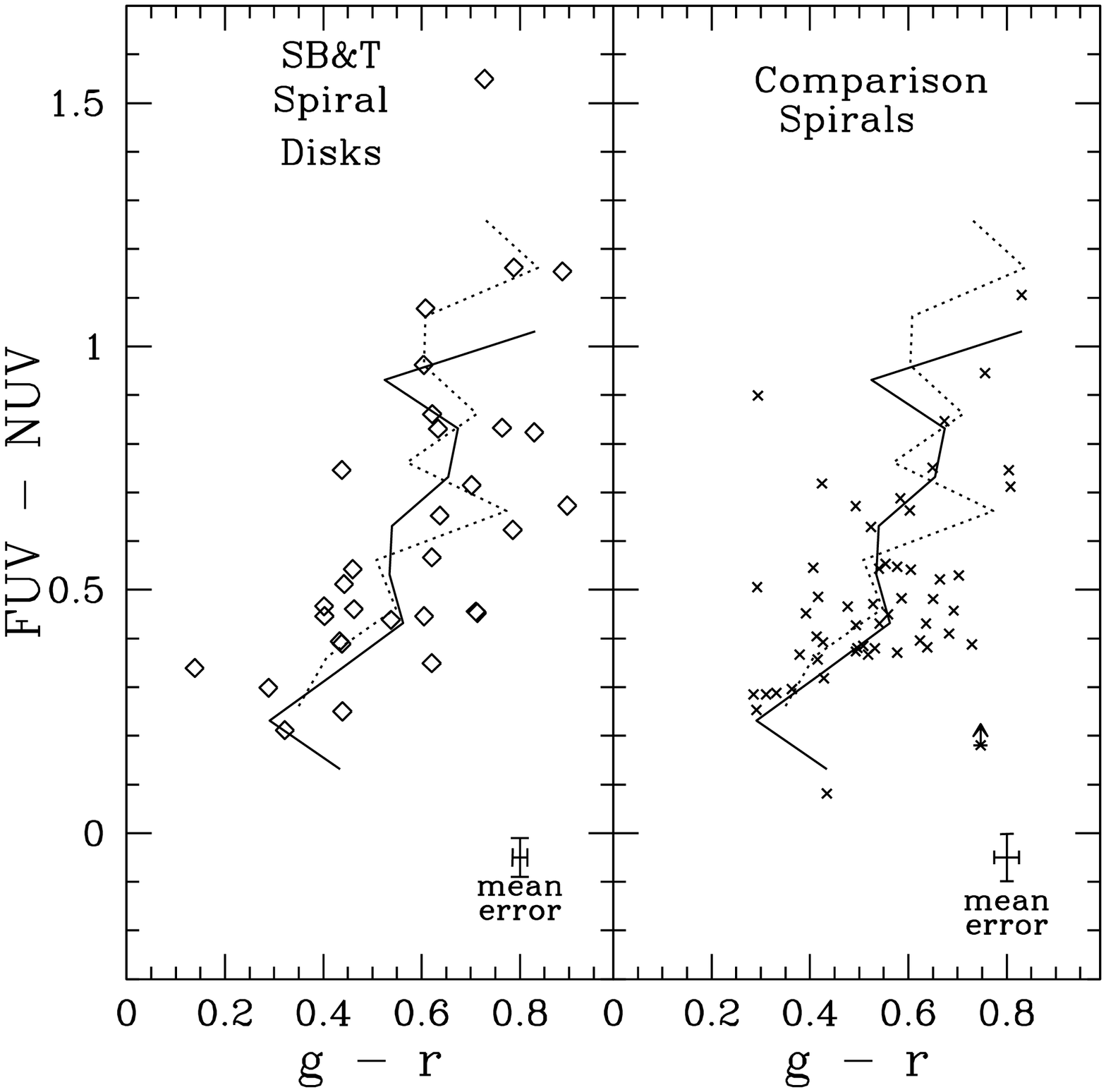}
\caption{
  \small 
The GALEX FUV $-$ NUV vs.\ g $-$ r colors for the
main disks of the interacting galaxy sample classified
by NED as S0/a $-$ Sd (left panel; open diamonds),
compared to the spirals in the full comparison sample
(right panel; crosses).
The solid line shows the mean values of g $-$ r for the full comparison
sample calculated in 0.1 magnitude bins of FUV $-$ NUV, 
while the dotted line gives the mean values for the 
SB\&T spirals.
The mean statistical uncertainties for the datapoints
are plotted in the lower left.
}
\end{figure}

\begin{figure}
\plotone{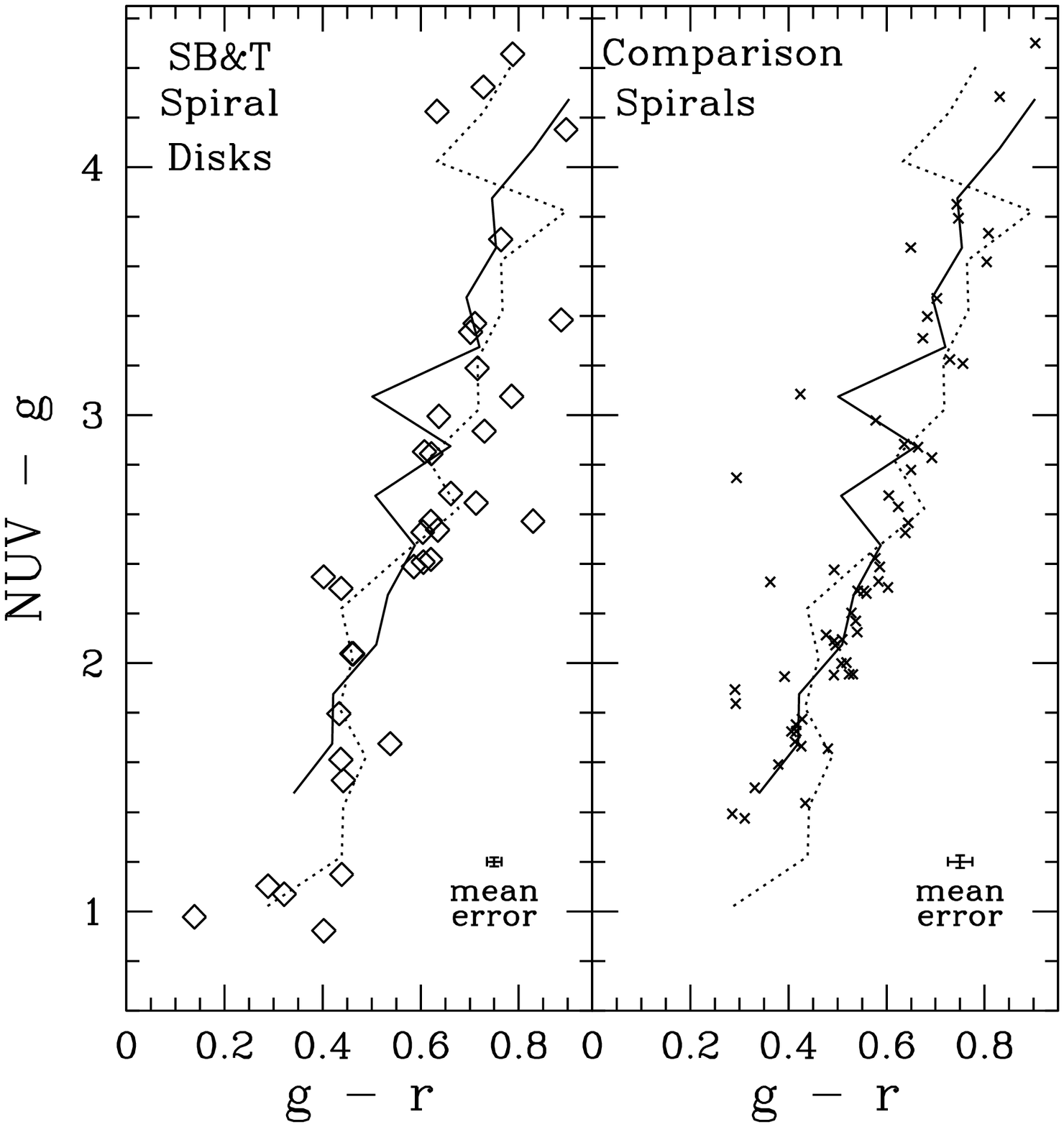}
\caption{
  \small 
The GALEX NUV $-$ g vs.\ g $-$ r colors for the
main disks of the interacting galaxy sample classified
by NED as S0/a $-$ Sd (left panel; open diamonds),
compared to the spirals in the full comparison sample
(right panel; crosses).
The solid line shows the mean values of g $-$ r for the full comparison
sample calculated in 0.2 magnitude bins of NUV $-$ g, 
while the dotted line gives the mean values for the 
SB\&T spirals.
The mean statistical uncertainties are plotted in the lower left.
}
\end{figure}

\begin{figure}
\plotone{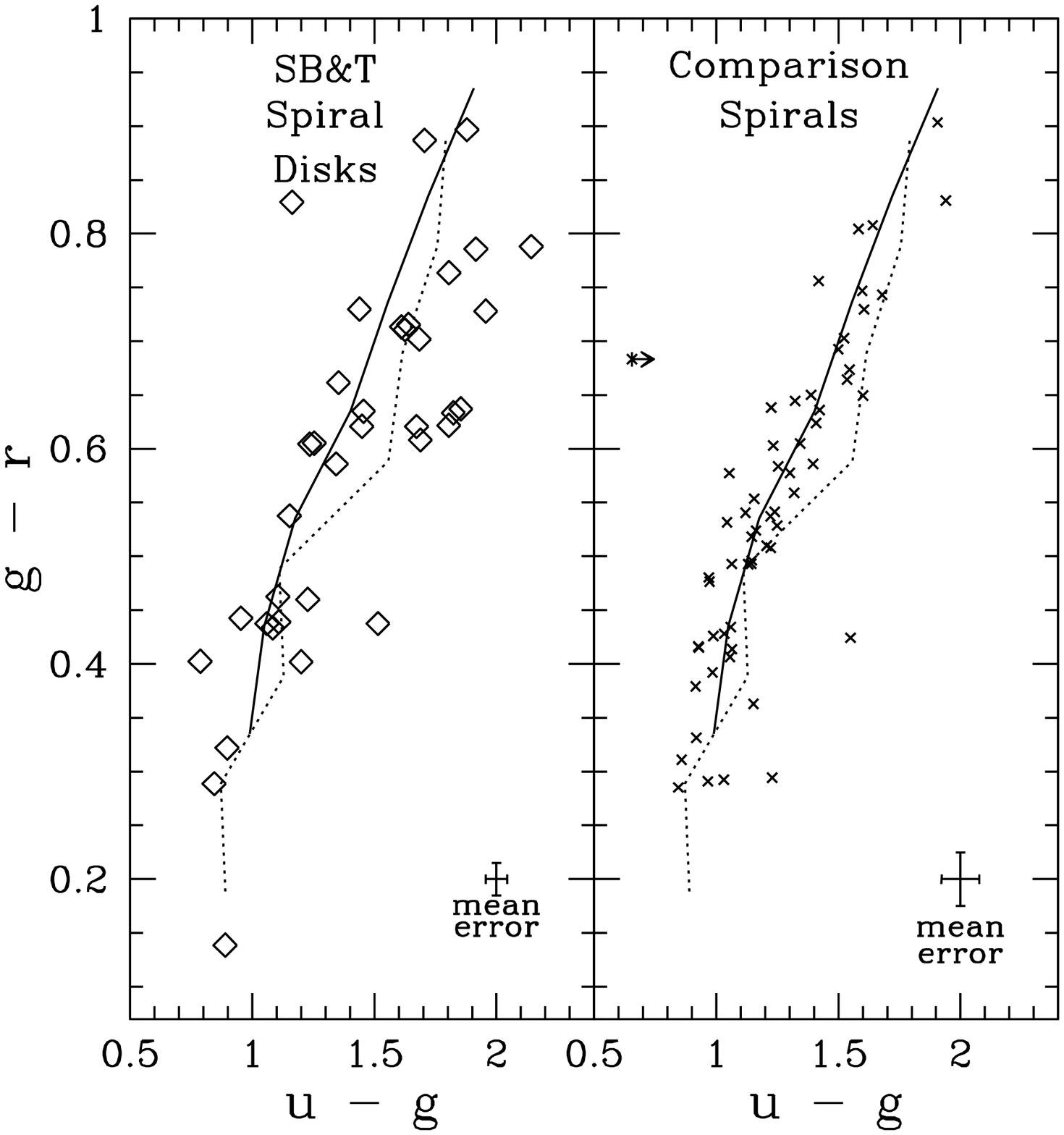}
\caption{
  \small 
The GALEX u $-$ g vs.\ g $-$ r colors for the
main disks of the interacting galaxy sample classified
by NED as S0/a $-$ Sd (open diamonds),
compared to the spirals in the full comparison sample
(crosses).
The solid line shows the mean values of u $-$ g for the comparison
sample calculated in 0.1 magnitude bins of g $-$ r, 
while the dotted line gives the mean values for the 
SB\&T spirals.
The mean statistical uncertainties are plotted in the lower left.
}
\end{figure}

\begin{figure}
\plotone{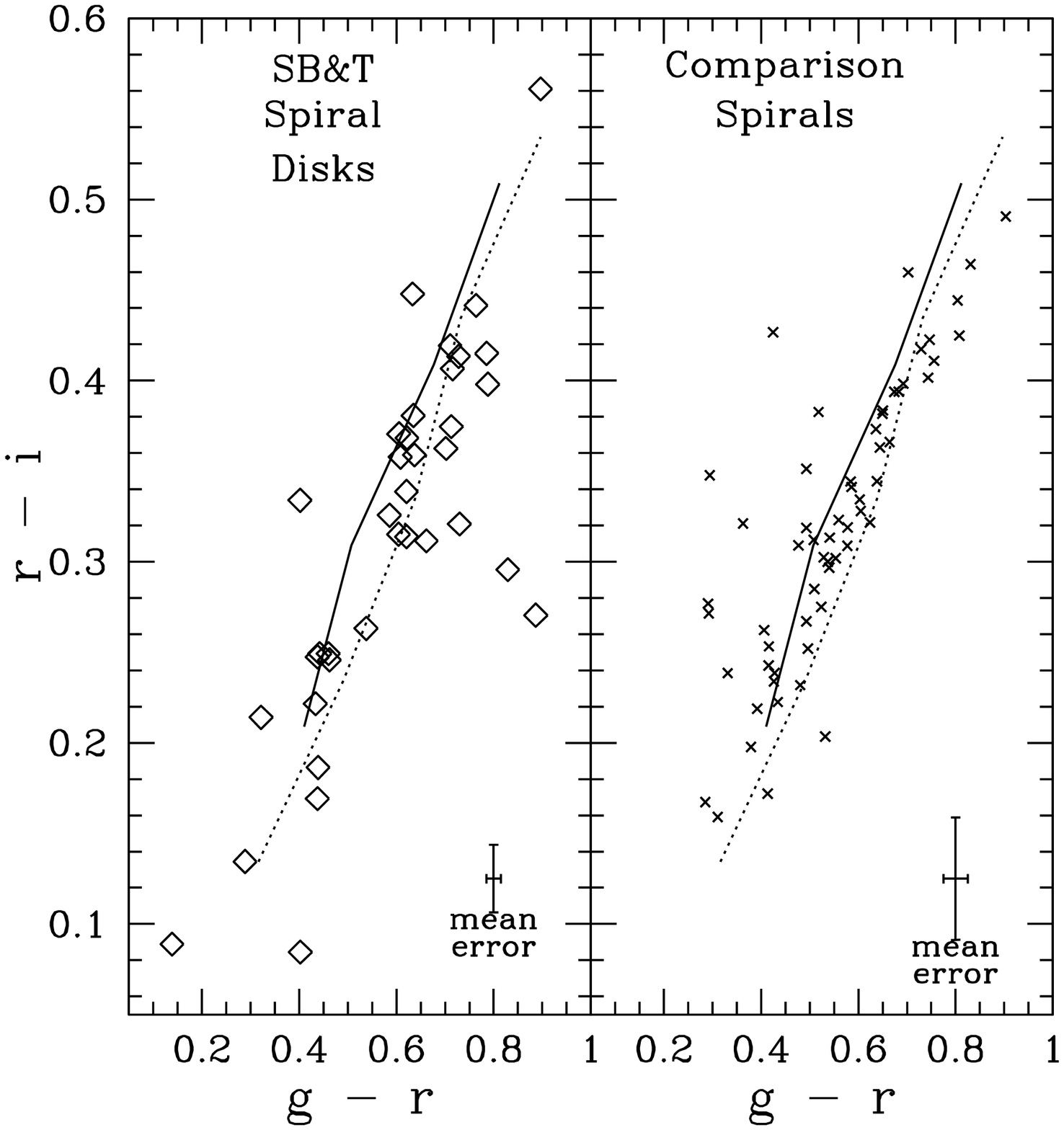}
\caption{
  \small 
The GALEX g $-$ r vs.\ r $-$ i colors for the
main disks of the interacting galaxy sample classified
by NED as S0/a $-$ Sd (open diamonds),
compared to the spirals in the full comparison sample
(crosses).
The solid line shows the mean values of g $-$ r for the comparison
sample calculated in 0.1 magnitude bins of r $-$ i, 
while the dotted line gives the mean values for the 
SB\&T spirals.
The mean statistical uncertainties are plotted in the lower left.
}
\end{figure}

\begin{figure}
\plotone{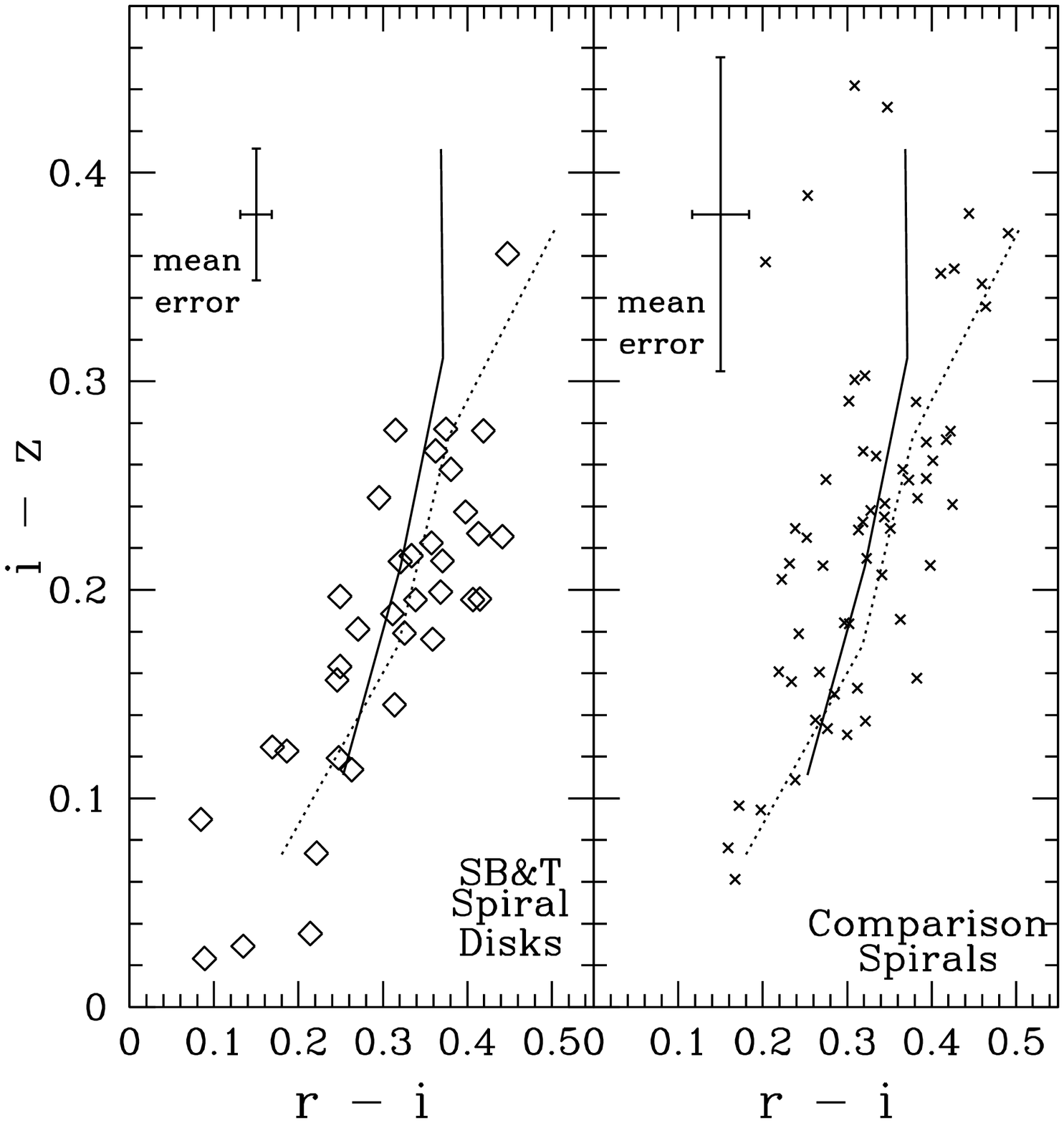}
\caption{
  \small 
The GALEX r $-$ i vs.\ i $-$ z colors for the
main disks of the interacting galaxy sample classified
by NED as S0/a $-$ Sd (open diamonds),
compared to the spirals in the full comparison sample
(crosses).
The solid line shows the mean values of r $-$ i for the comparison
sample calculated in 0.1 magnitude bins of i $-$ z, 
while the dotted line gives the mean values for the 
SB\&T spirals.
The mean statistical uncertainties are plotted in the lower left.
}
\end{figure}

\begin{figure}
\plotone{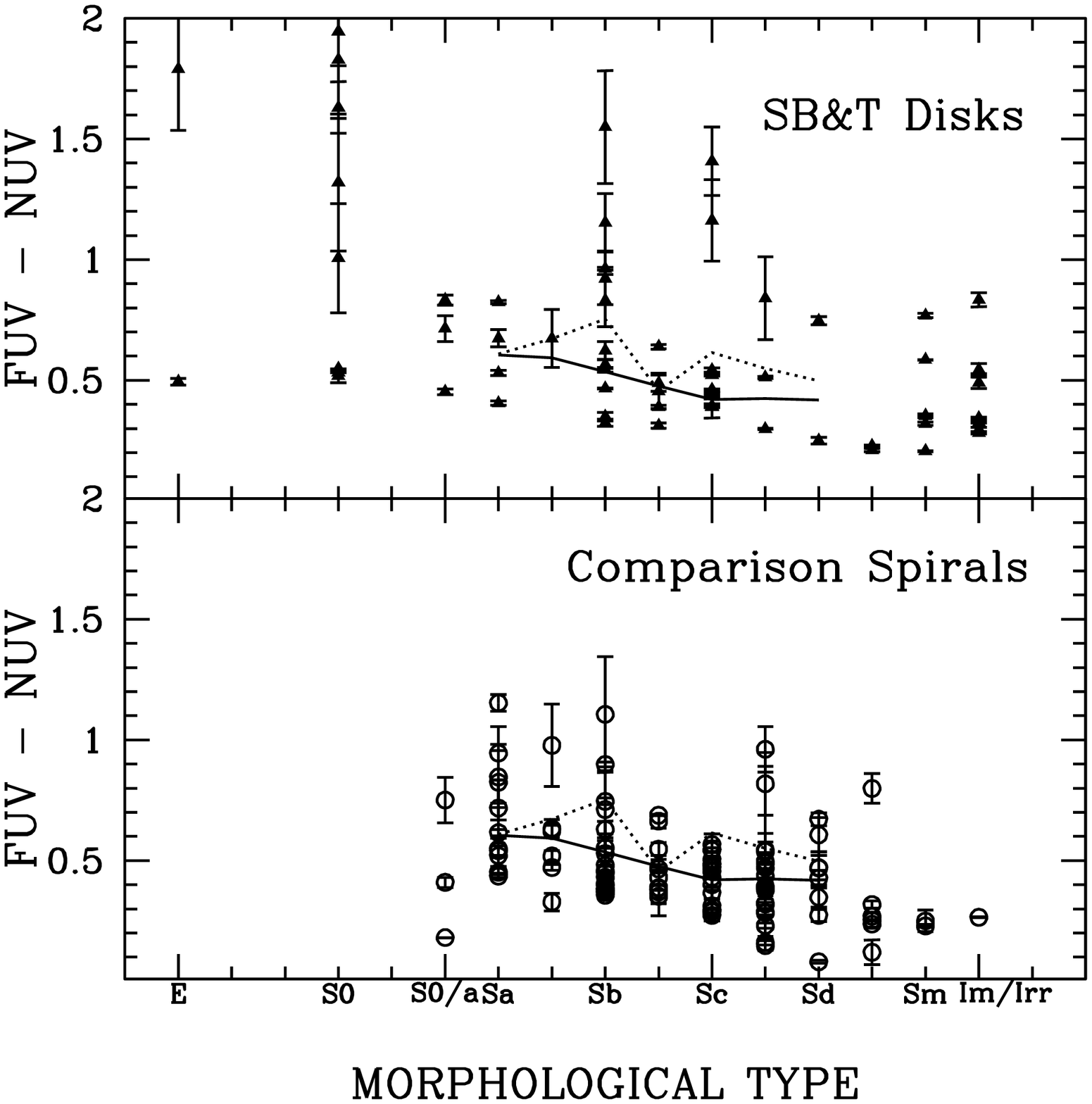}
\caption{
  \small 
Plots of FUV $-$ NUV color vs.\ morphological type for
the SB\&T sample and the full comparison sample.
The solid lines show the mean-value line for types Sa $-$ Sd for the
comparison spirals, while the dotted line marks the mean-value curve
for the interacting sample.
Statistical uncertainties are shown.
}
\end{figure}
\clearpage

\begin{figure}
\plotone{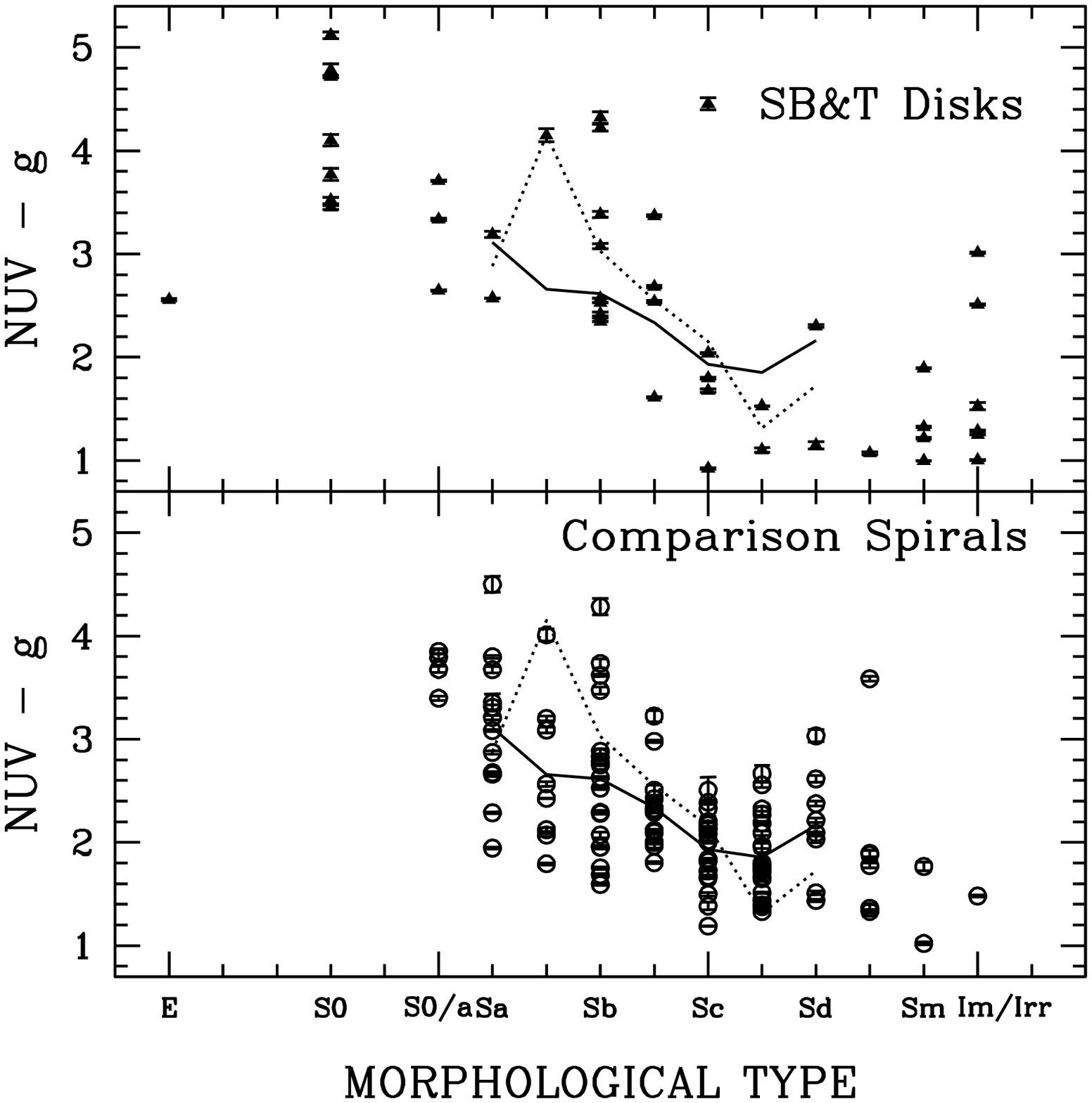}
\caption{
  \small 
Plots of NUV $-$ g color vs.\ morphological type for
the SB\&T sample and the full comparison sample.
The solid lines show the mean-value line for types Sa $-$ Sd for the
comparison spirals, while the dotted line marks the mean-value curve
for the interacting sample.
Statistical uncertainties are shown.
}
\end{figure}

\begin{figure}
\plotone{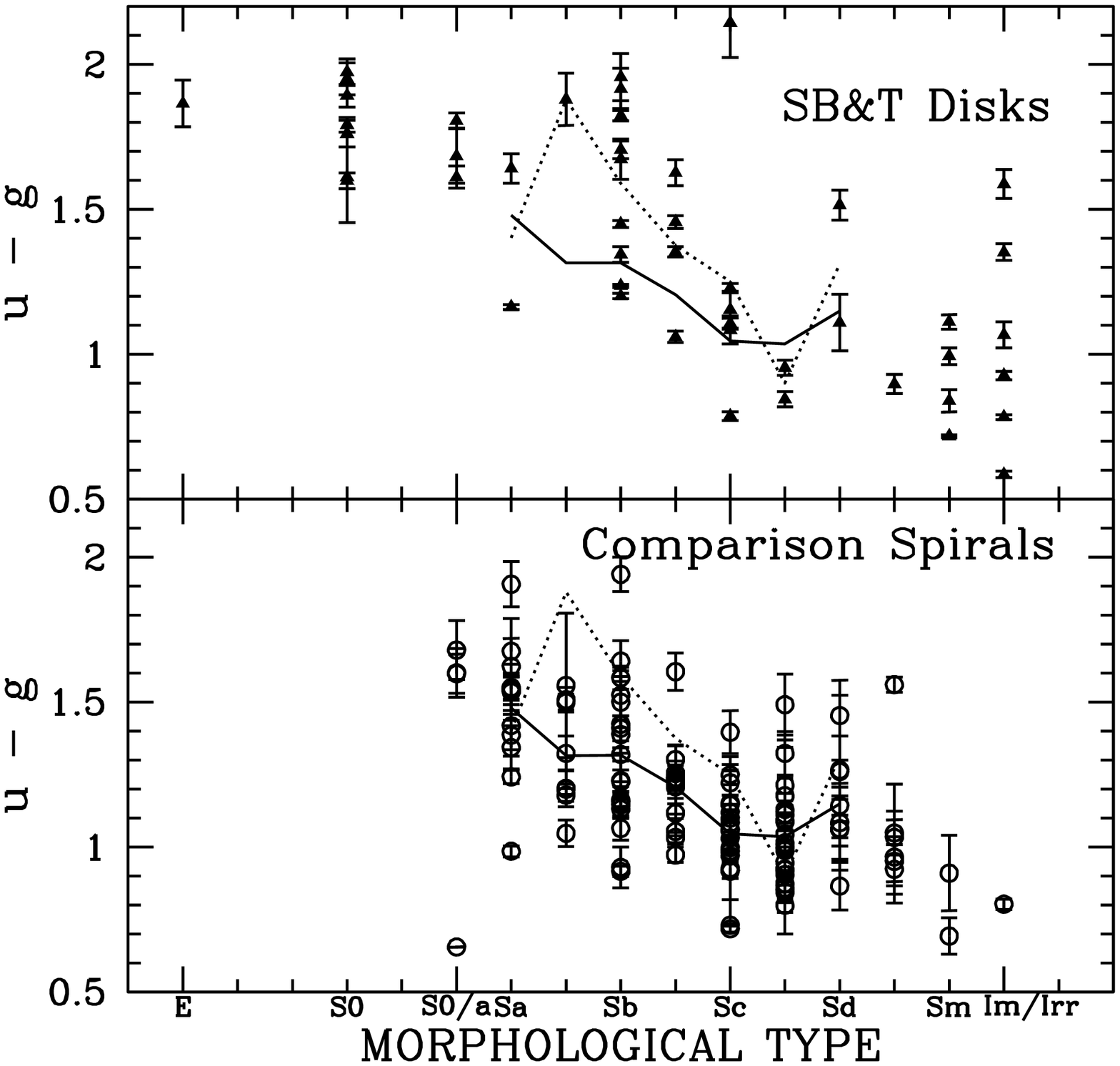}
\caption{
  \small 
Plots of u $-$ g color vs.\ morphological type for
the SB\&T sample and the full comparison sample.
The solid lines show the mean-value line for types Sa $-$ Sd for the
comparison spirals, while the dotted line marks the mean-value curve
for the interacting sample.
Statistical uncertainties are shown.
}
\end{figure}

\begin{figure}
\plotone{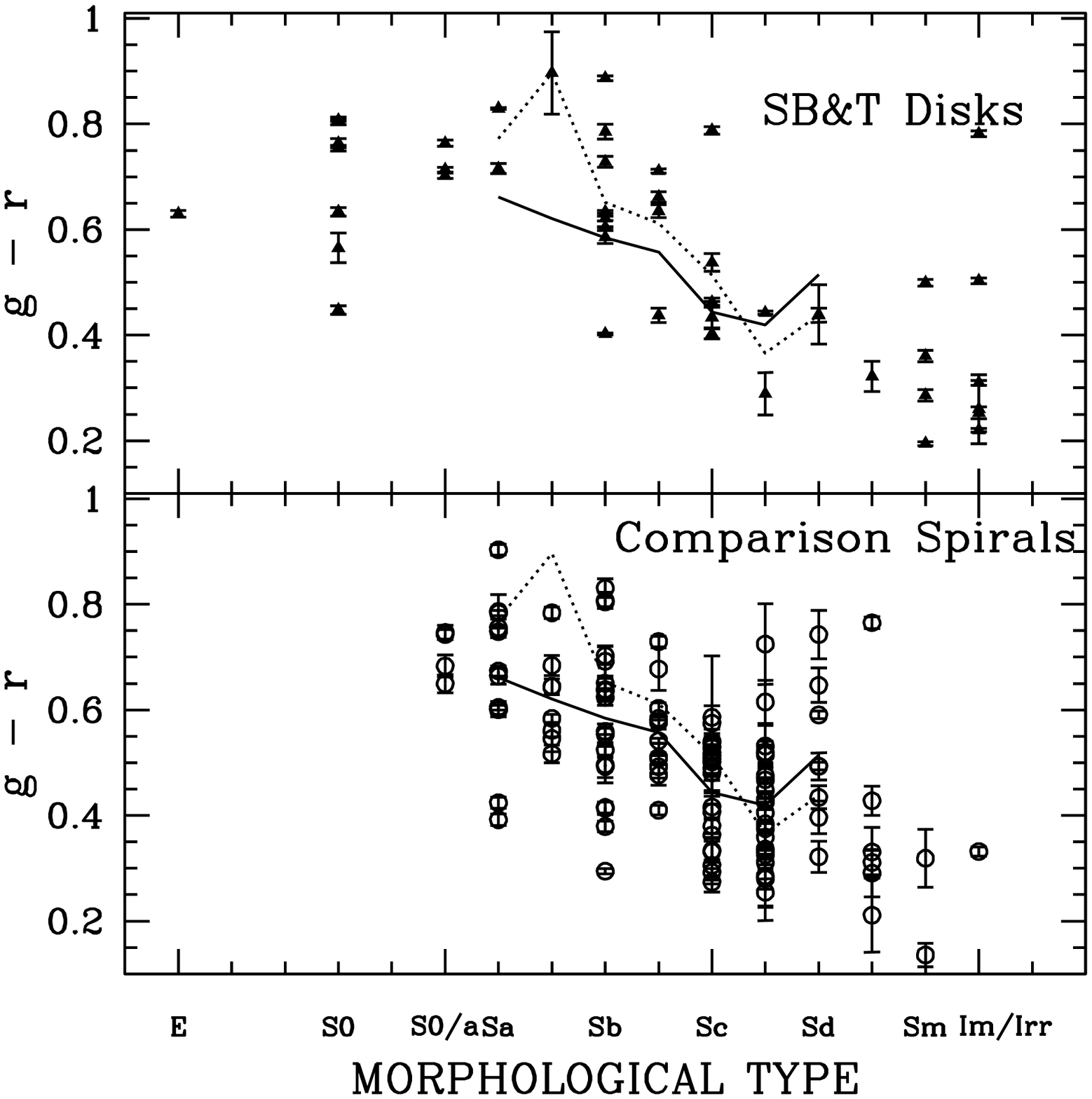}
\caption{
  \small 
Plots of g $-$ r color vs.\ morphological type for
the SB\&T sample and the full comparison sample.
The solid lines show the mean-value line for types Sa $-$ Sd for the
comparison spirals, while the dotted line marks the mean-value curve
for the interacting sample.
Statistical uncertainties are shown.
}
\end{figure}

\begin{figure}
\plotone{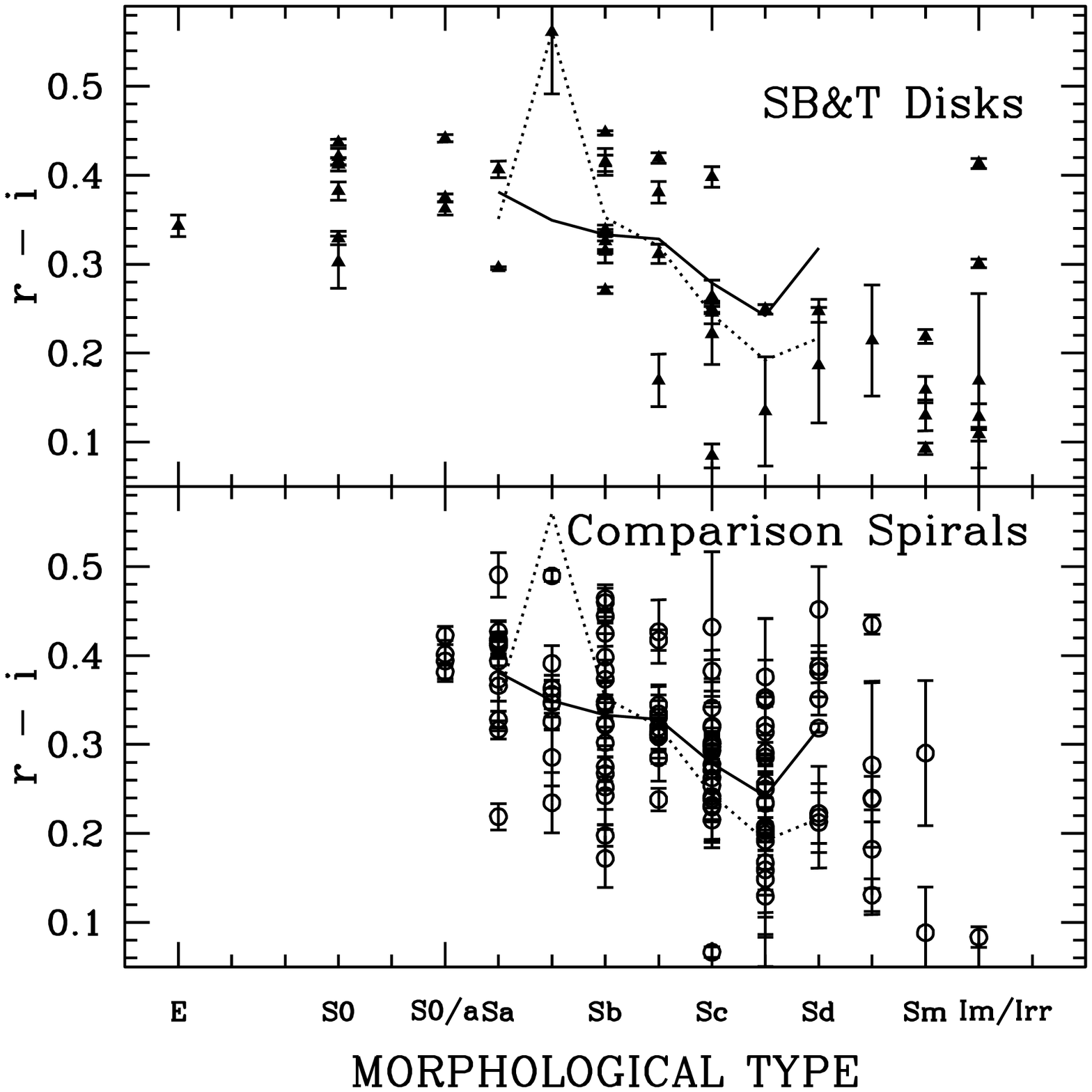}
\caption{
  \small 
Plots of r $-$ i color vs.\ morphological type for
the SB\&T sample and the full comparison sample.
The solid lines show the mean-value line for types Sa $-$ Sd for the
comparison spirals, while the dotted line marks the mean-value relation
for the interacting sample.
Statistical uncertainties are shown.
}
\end{figure}

\begin{figure}
\plotone{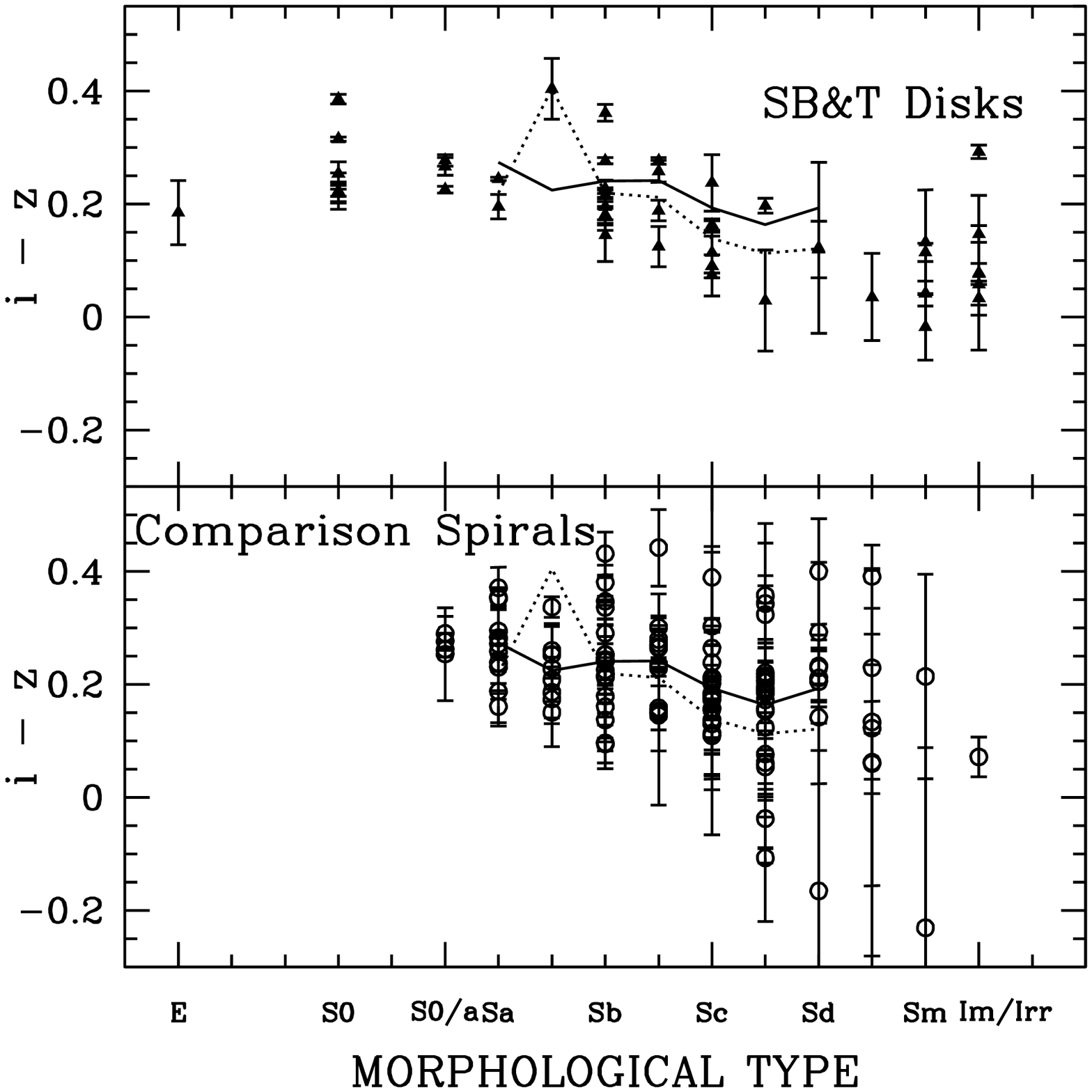}
\caption{
  \small 
Plots of i $-$ z color vs.\ morphological type for
the SB\&T sample and the full comparison sample.
The solid lines show the mean-value line for types Sa $-$ Sd for the
comparison spirals, while the dotted line marks the mean-value relation
for the interacting sample.
Statistical uncertainties are shown.
}
\end{figure}

\begin{figure}
\plotone{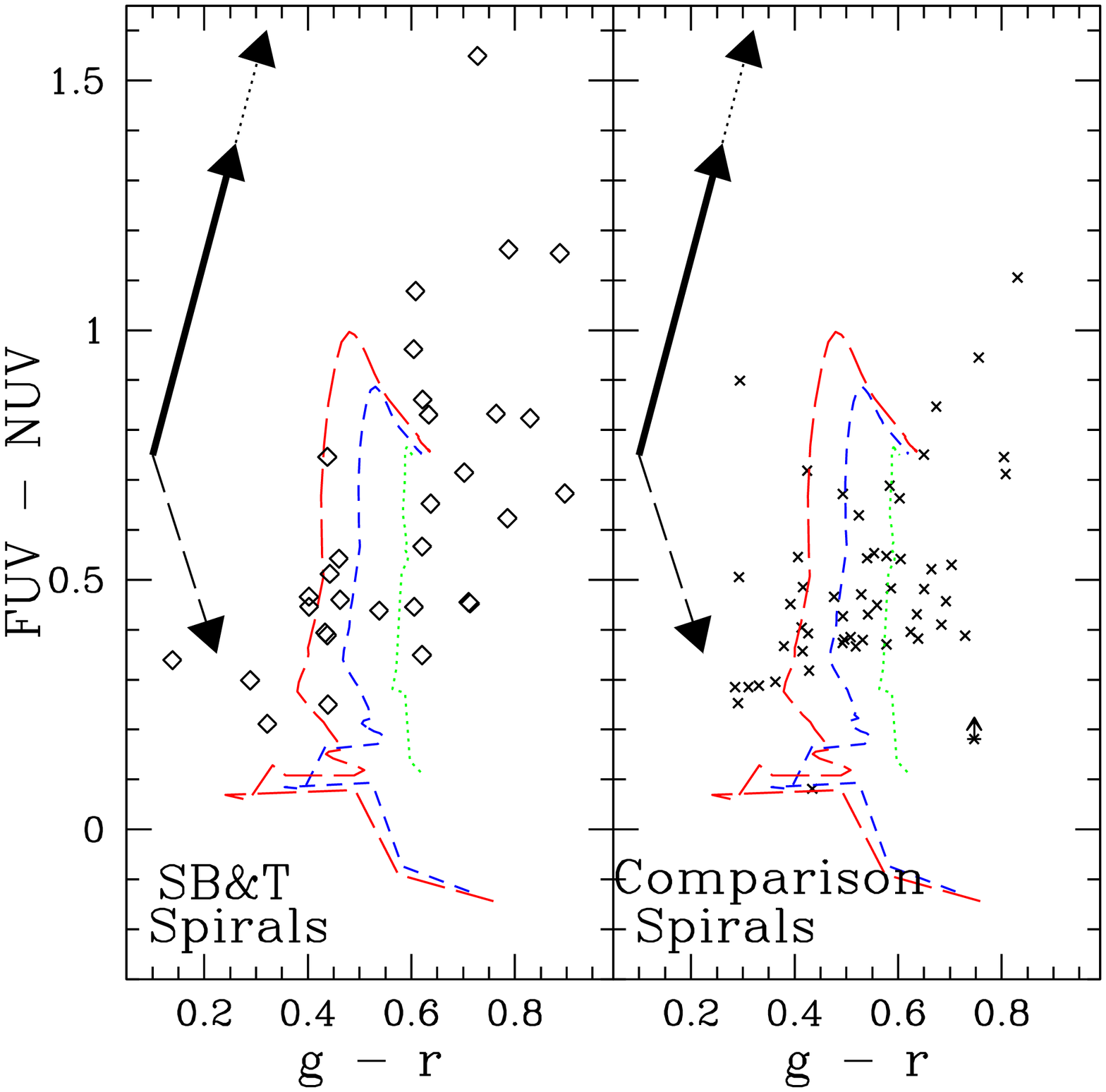}
\caption{
  \small 
The GALEX FUV $-$ NUV vs.\ g $-$ r colors for the
main disks of the interacting galaxy sample classified
by NED as S0/a $-$ Sd (left panel; open diamonds),
compared to the spirals in the comparison sample
(right panel; crosses).
Solar metallicity zero-extinction synthesis
models are superimposed, consisting
of a typical spiral galaxy (see text)
plus an instantaneous burst.
The dotted
and dashed curves mark models of constant burst strength but increasing age,
starting with 10$^6$ years at the bottom of the page.
Three 
burst strengths are plotted, with
f$_z$(young)/f$_z$(old) = 0.01 (green dotted),
0.1 (blue short dashed), and 0.2 (red long dashes).
All models include H$\alpha$ but no other emission lines.
The sharp bend in the models to red g $-$ r colors at young ages is due to 
contributions from H$\alpha$.
The three arrows show reddening vectors as described in the text.
}
\end{figure}

\clearpage

\begin{figure}
\plotone{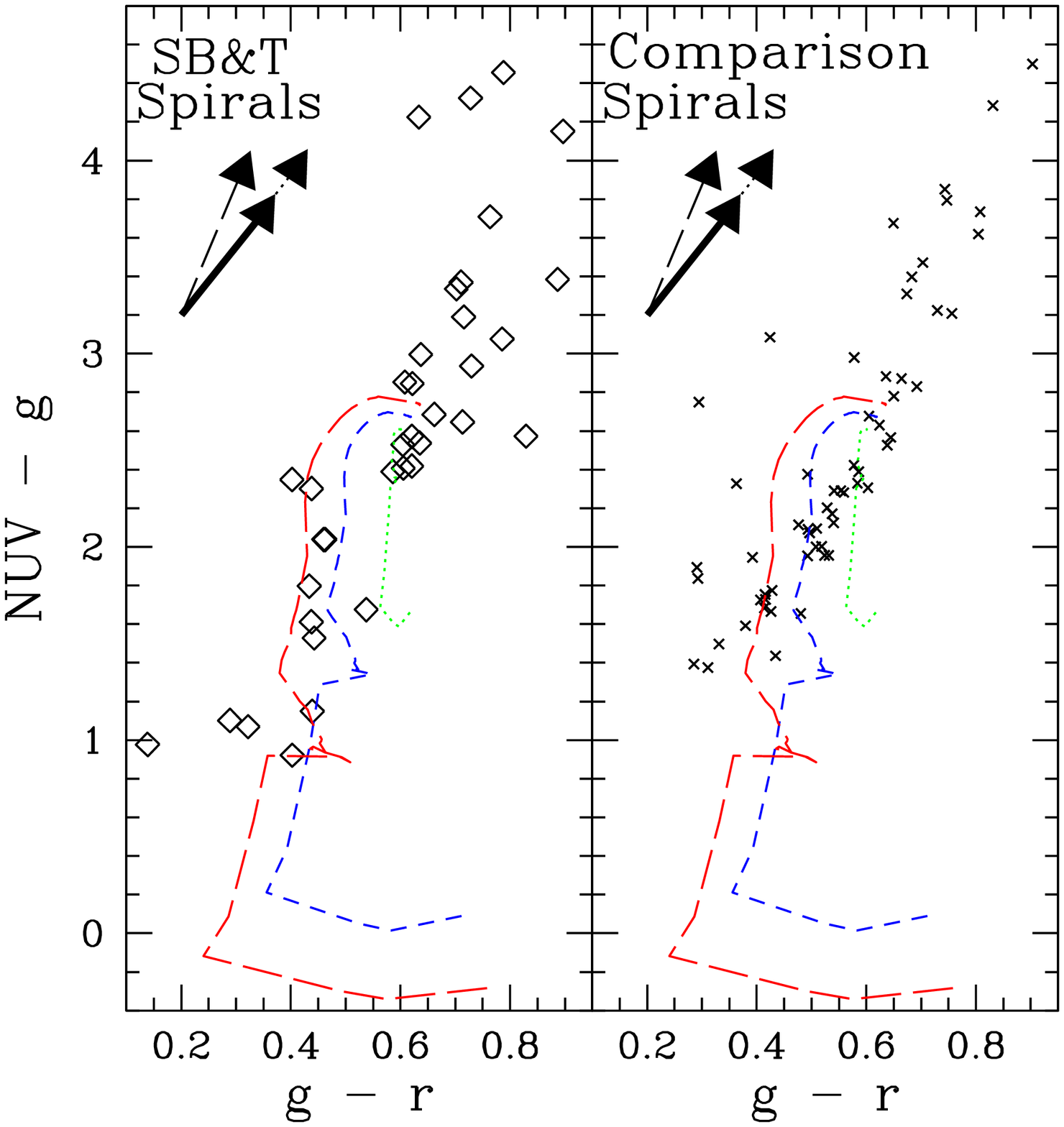}
\caption{
  \small 
The GALEX NUV $-$ g vs.\ g $-$ r colors for the
main disks of the interacting galaxy sample classified
by NED as S0/a $-$ Sd (left panel; open diamonds),
compared to the spirals in the comparison sample
(right panel; crosses).  
Solar metallicity zero-extinction synthesis
models are superimposed, consisting
of a typical spiral galaxy (see text)
plus an instantaneous burst.
The dotted
and dashed curves mark models of constant burst strength but increasing age,
starting with 10$^6$ years at the bottom of the page.
Three
burst strengths are plotted, with
f$_z$(young)/f$_z$(old) of 0.01 (green dotted),
0.1 (blue short dashed), and 0.2 (red long dashes).
All models include H$\alpha$ but no other emission lines.
The three arrows show reddening vectors as described in the text.
}
\end{figure}

\clearpage

\begin{figure}
\plotone{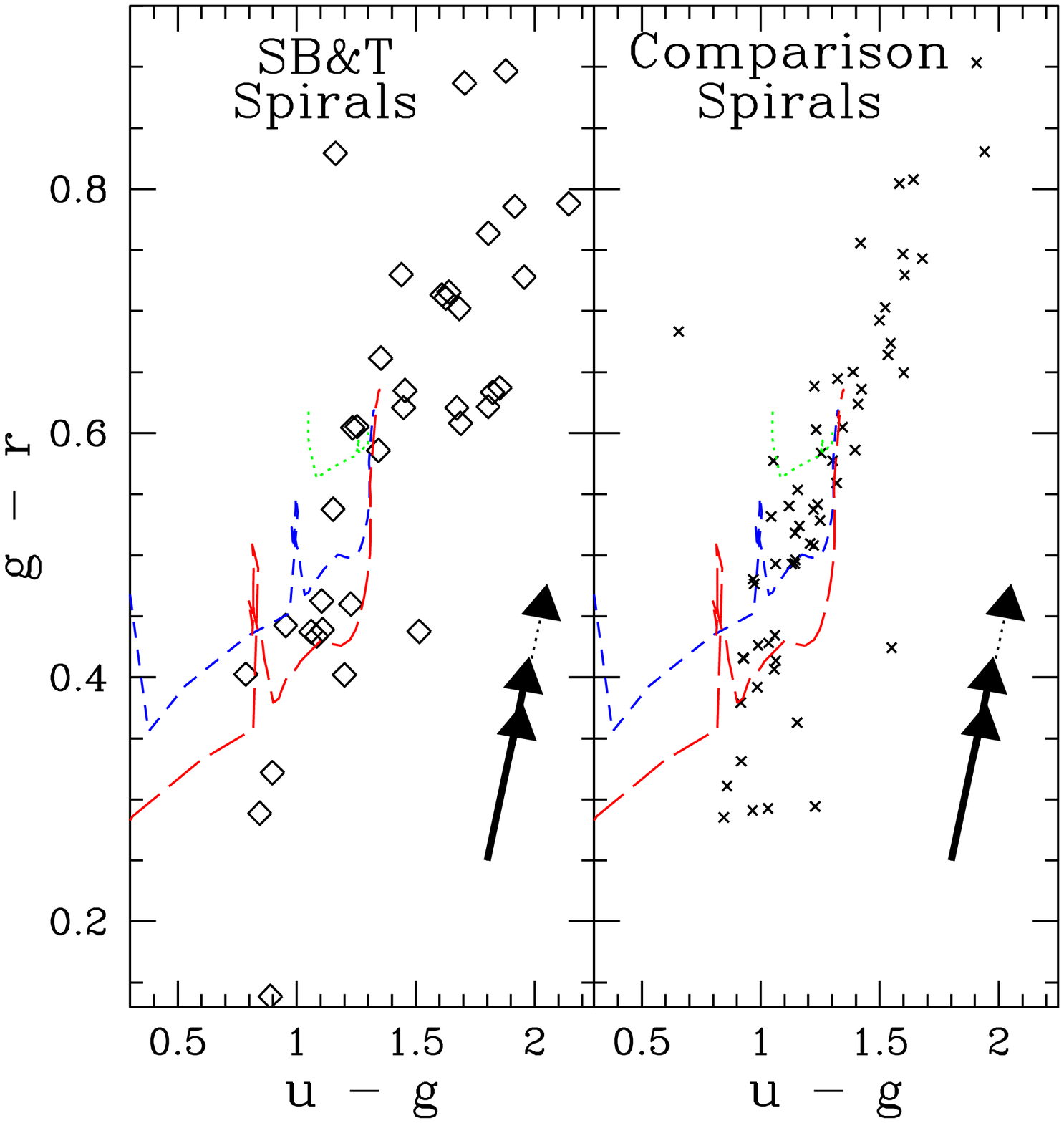}
\caption{
  \small 
The GALEX u $-$ g vs.\ g $-$ r colors for the
main disks of the interacting galaxy sample classified
by NED as S0/a $-$ Sd (open diamonds),
compared to the spirals in the comparison sample
(crosses).
Solar metallicity zero-extinction synthesis
models are superimposed, consisting
of a typical spiral galaxy (see text)
plus an instantaneous burst.
The dotted
and dashed curves mark models of constant burst strength but increasing age,
starting with 10$^6$ years at the bottom of the page.
Three burst strengths are plotted,
with f$_z$(young)/f$_z$(old) of 0.01 (green dotted),
0.1 (blue short dashed), and 0.2 (red long dashes).
All models include H$\alpha$ but no other emission lines.
The three arrows show reddening vectors as described in the text.
}
\end{figure}

\begin{figure}
\plotone{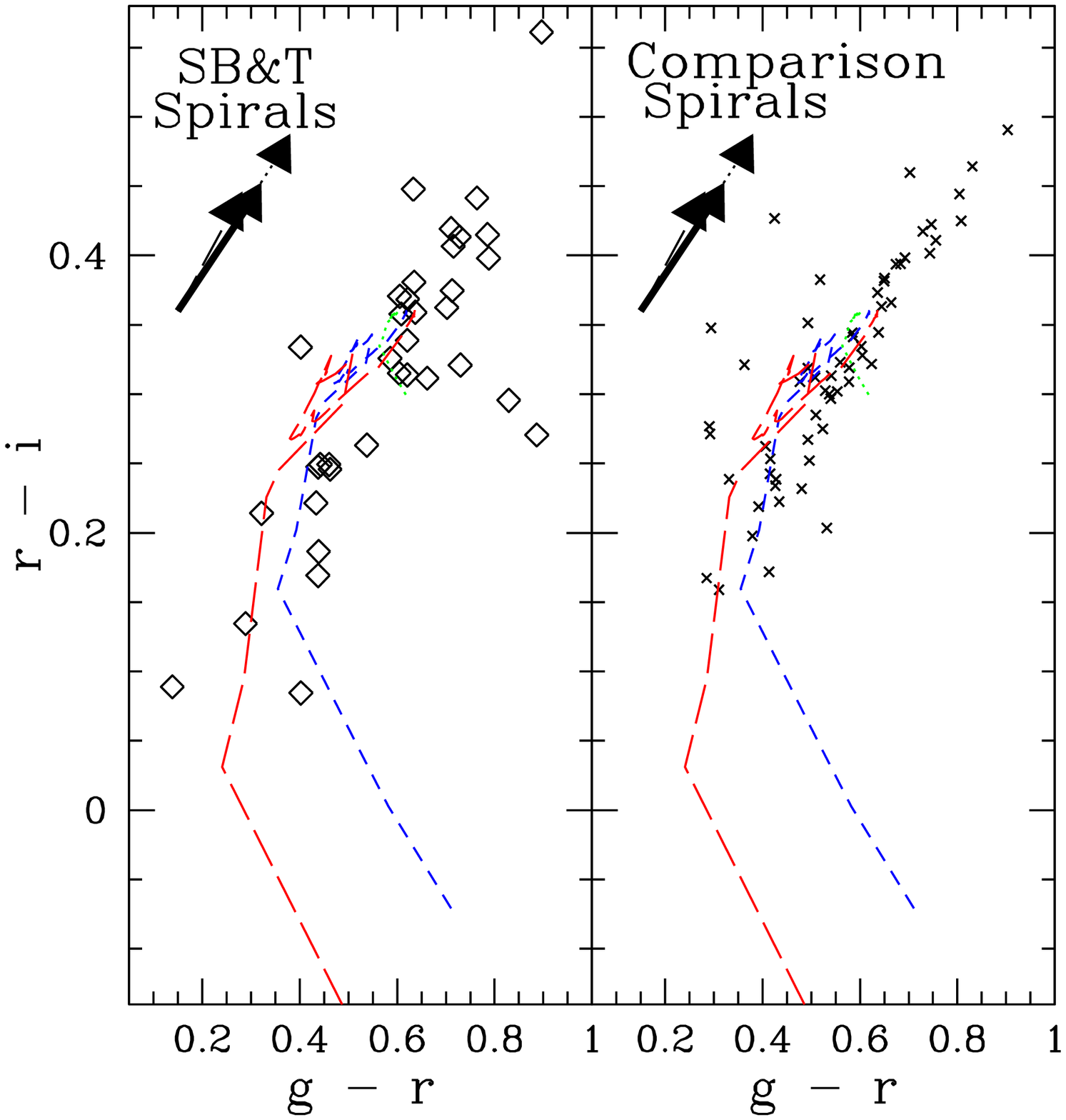}
\caption{
  \small 
The GALEX g $-$ r vs.\ r $-$ i colors for the
main disks of the interacting galaxy sample classified
by NED as S0/a $-$ Sd (open diamonds),
compared to the spirals in the comparison sample
(crosses).
Solar metallicity zero-extinction synthesis
models are superimposed, consisting
of a typical spiral galaxy (see text)
plus an instantaneous burst.
The dotted
and dashed curves mark models of constant burst strength but increasing age,
starting with 10$^6$ years at the bottom of the page.
Three burst strengths are plotted, with
f$_z$(young)/f$_z$(old) of 0.01 (green dotted),
0.1 (blue short dashed), and 0.2 (red long dashes).
The three arrows show reddening vectors as described in the text.
}
\end{figure}

\begin{figure}
\plotone{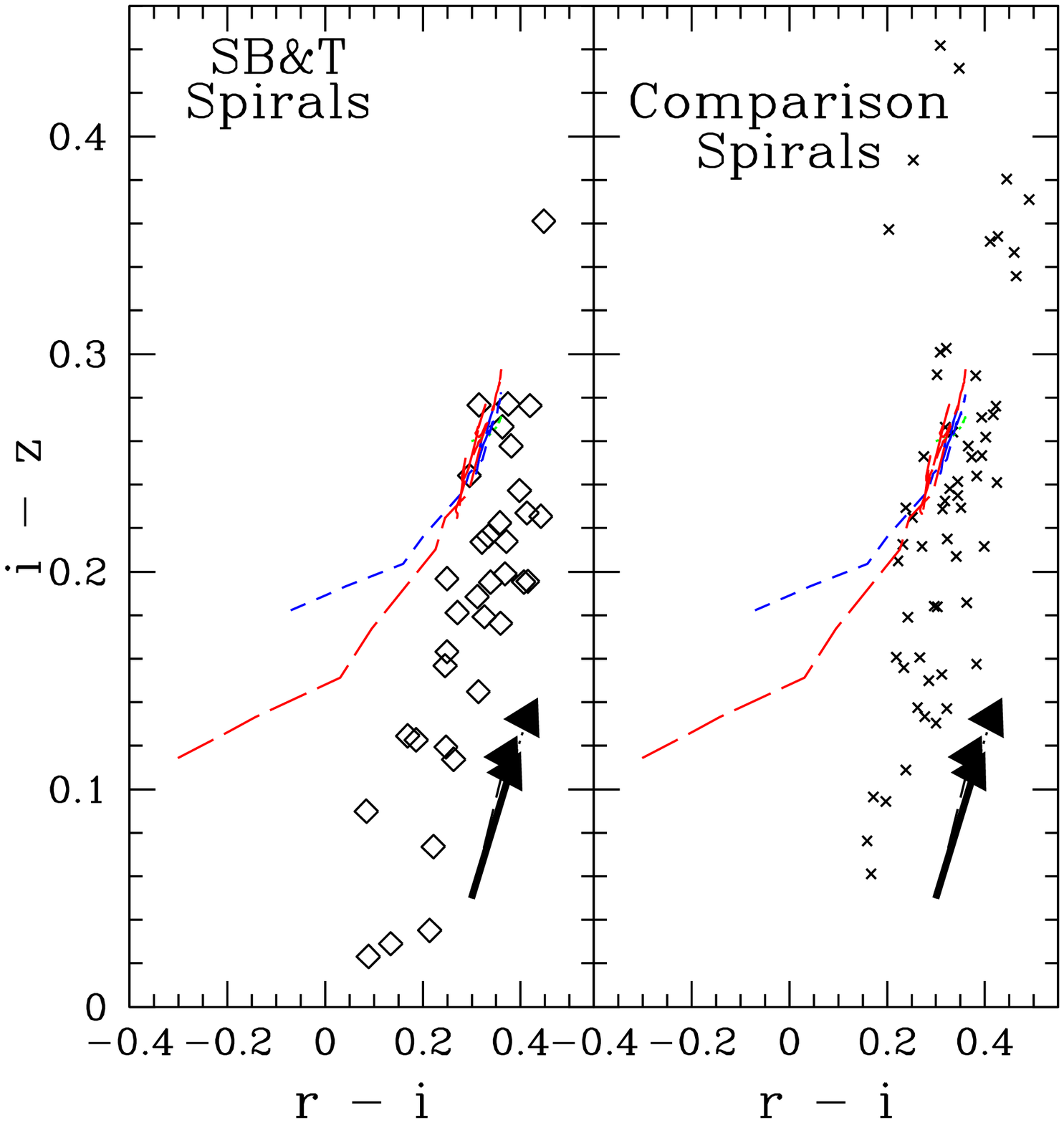}
\caption{
  \small 
The GALEX r $-$ i vs.\ i $-$ z colors for the
main disks of the interacting galaxy sample classified
by NED as S0/a $-$ Sd (open diamonds),
compared to the spirals in the comparison sample
(crosses).
Solar metallicity zero-extinction synthesis
models are superimposed, consisting
of a typical spiral galaxy (see text)
plus an instantaneous burst.
The dotted
and dashed curves mark models of constant burst strength but increasing age,
starting with 10$^6$ years at the bottom of the page.
Three 
burst strengths are plotted, with
f$_z$(young)/f$_z$(old) of 0.01 (green dotted),
0.1 (blue short dashed), and 0.2 (red long dashes).
All models include H$\alpha$ but no other emission lines.
At these wavelengths, the three reddening vectors are very similar.
}
\end{figure}

\begin{deluxetable}{cccccccc}
\tabletypesize{\scriptsize}
\setlength{\tabcolsep}{0.03in}
\def\et#1#2#3{${#1}^{+#2}_{-#3}$}
\tablewidth{0pt}
\tablecaption{Optical and UV Magnitudes for Normal Spiral Galaxy Sample$^a$\label{tab-4}}
\tablehead{
\multicolumn{1}{c}{} &
\colhead{FUV} & 
\colhead{NUV} & 
\colhead{u} & 
\colhead{g} & 
\colhead{r} & 
\colhead{i} & 
\colhead{z} 
\\ 
\multicolumn{1}{c}{Name} &
\colhead{(mag)} & 
\colhead{(mag)} & 
\colhead{(mag)} & 
\colhead{(mag)} & 
\colhead{(mag)} & 
\colhead{(mag)} & 
\colhead{(mag)} 
\\ 
}
\startdata
CGCG 0377-039 &     17.86 $\pm$    0.09 &     17.34 $\pm$    0.01 &     15.91 $\pm$    0.05 &     14.49 $\pm$    0.01 &     13.85 $\pm$    0.01 &     13.47 $\pm$    0.01 &     13.26 $\pm$    0.01   \\
     IC 0159 &     15.89 $\pm$    0.00 &     15.48 $\pm$    0.00 &     14.86 $\pm$    0.04 &     13.80 $\pm$    0.01 &     13.39 $\pm$    0.01 &     13.21 $\pm$    0.03 &     13.14 $\pm$    0.03   \\
     IC 0653 &$\ge$17.38               &     17.20 $\pm$    0.03 &     15.01 $\pm$    0.07 &     13.41 $\pm$    0.01 &     12.66 $\pm$    0.01 &     12.24 $\pm$    0.01 &     12.00 $\pm$    0.01   \\
     IC 0673 &     16.35 $\pm$    0.03 &     15.92 $\pm$    0.02 &     14.99 $\pm$    0.16 &     13.85 $\pm$    0.03 &     13.34 $\pm$    0.03 &     13.04 $\pm$    0.05 &     12.89 $\pm$    0.13   \\
     IC 0716 &     18.02 $\pm$    0.10 &     17.63 $\pm$    0.05 &     16.02 $\pm$    0.06 &     14.41 $\pm$    0.01 &     13.68 $\pm$    0.01 &     13.26 $\pm$    0.01 &     13.01 $\pm$    0.04   \\
     IC 0952 &     16.88 $\pm$    0.05 &     16.45 $\pm$    0.00 &     15.40 $\pm$    0.04 &     14.16 $\pm$    0.00 &     13.62 $\pm$    0.00 &     13.31 $\pm$    0.01 &     13.10 $\pm$    0.01   \\
     IC 1221 &     15.38 $\pm$    0.00 &     15.30 $\pm$    0.00 &     14.93 $\pm$    0.03 &     13.87 $\pm$    0.01 &     13.43 $\pm$    0.02 &     13.21 $\pm$    0.03 &     13.02 $\pm$    0.07   \\
     IC 1222 &     16.23 $\pm$    0.03 &     15.76 $\pm$    0.00 &     14.81 $\pm$    0.03 &     13.56 $\pm$    0.00 &     13.03 $\pm$    0.00 &     12.73 $\pm$    0.01 &     12.56 $\pm$    0.02   \\
     IC 3467 &     17.54 $\pm$    0.35 &     17.00 $\pm$    0.02 &     16.32 $\pm$    0.12 &     15.20 $\pm$    0.03 &     14.82 $\pm$    0.04 &     14.53 $\pm$    0.03 &     14.36 $\pm$    0.07   \\
     IC 4229 &     16.46 $\pm$    0.03 &     15.83 $\pm$    0.00 &     15.04 $\pm$    0.02 &     13.88 $\pm$    0.01 &     13.35 $\pm$    0.01 &     13.08 $\pm$    0.01 &     12.85 $\pm$    0.02   \\
\enddata
\tablenotetext{a}{This table is available in its entirety in the electronic version of the journal. 
A shortened version is presented here to illustrate format and content.  The quoted uncertainties include only statistical errors.}
\end{deluxetable}


%
%
\begin{deluxetable}{|c|c |c | c|c|}
\tabletypesize{\scriptsize}
\def\et#1#2#3{${#1}^{+#2}_{-#3}$}
\tablewidth{0pt}
\tablecaption{RMS Deviations
for the Color-Color Plots$^a$\label{tab-2}}
\tablehead{
\multicolumn{1}{c}{Colors} &
\multicolumn{1}{c}{Full} &
\multicolumn{1}{c}{Mean} &
\multicolumn{1}{c}{SB\&T } &
\multicolumn{1}{c}{SB\&T } 
\\
\multicolumn{1}{c}{} &
\multicolumn{1}{c}{ Comparison } &
\multicolumn{1}{c}{ Type-Matched} &
\multicolumn{1}{c}{Spirals} &
\multicolumn{1}{c}{Spirals} \\
\multicolumn{1}{c}{} &
\multicolumn{1}{c}{Sample} &
\multicolumn{1}{c}{Subsets} &
\multicolumn{1}{c}{vs.} &
\multicolumn{1}{c}{vs.} \\
\multicolumn{1}{c}{} &
\multicolumn{1}{c}{} &
\multicolumn{1}{c}{} &
\multicolumn{1}{c}{own mean} &
\multicolumn{1}{c}{comp. mean} \\
}
\startdata
FUV $-$ NUV vs. g $-$ r &
0.111
 & 
0.108
 & 
0.128
 & 
0.156
 \\ 
NUV $-$ g vs. g $-$ r & 
0.065
 & 
0.060
 & 
0.079
 & 
0.123
 \\ 
g $-$ r vs. u $-$ g & 
0.124
 & 
0.120
 & 
0.216
 & 
0.286
 \\ 
r $-$ i vs. g $-$ r & 
0.090
 & 
0.087
 & 
0.106
 & 
0.117
 \\ 
i $-$ z vs. r $-$ i & 
0.076
 & 
0.063
 & 
0.062
 & 
0.071
 \\ 
\enddata
\tablenotetext{a}
{Upper/lower limits are not included in
these statistics.
}
\end{deluxetable}
%


%
%
\begin{deluxetable}{|c|c | c | c|}
\tabletypesize{\scriptsize}
\def\et#1#2#3{${#1}^{+#2}_{-#3}$}
\tablewidth{0pt}
\tablecaption{RMS Deviations 
for Color vs. Type Plots$^a$\label{tab-3}}
\tablehead{
\multicolumn{1}{c}{Color} &
\multicolumn{1}{c}{Full } &
\multicolumn{1}{c}{Mean} &
\multicolumn{1}{c}{SB\&T } \\
\multicolumn{1}{c}{} &
\multicolumn{1}{c}{ Comparison } &
\multicolumn{1}{c}{ Type-Matched} &
\multicolumn{1}{c}{Spiral} \\
\multicolumn{1}{c}{} &
\multicolumn{1}{c}{ Sample } &
\multicolumn{1}{c}{ Subsets} &
\multicolumn{1}{c}{Sample$^b$} \\
}
\startdata
FUV $-$ NUV &
0.200
 & 
0.194
 & 
0.302
 \\ 
NUV $-$ g  & 
0.527
 & 
0.557
 & 
0.811
 \\ 
u $-$ g & 
0.194
 & 
0.201
 & 
0.310
 \\ 
g $-$ r & 
0.120
 & 
0.123
 & 
0.118
 \\ 
r $-$ i  & 
0.074
 & 
0.073
 & 
0.079
 \\ 
i $-$ z & 
0.090
 & 
0.081
 & 
0.061
 \\ 
\enddata
\tablenotetext{a}
{Upper/lower limits are not included in
these statistics.
}
\tablenotetext{b}
{Calculated relative to own mean curve.
}
\end{deluxetable}
%









\begin{thebibliography}{}

\bibitem[Abazajian et al.(2003)]{abazajian03}
Abazajian, K., et al.\ 2003, AJ, 126, 2081

\bibitem[Abraham et al.(1996)]{abraham96}
Abraham, R. G., van den Bergh, S., Glazebrook, K., Ellis, R. S.,
Santiago, B. X., Surma, P., \& Griffiths, R. E. 1996, ApJS, 107, 1

\bibitem[Arp(1966)]{arp66}
Arp, H. 1966, Atlas of Peculiar Galaxies (Pasadena: Caltech)


\bibitem[Barton et al.(2000)]{barton00}
Barton, E. J., Geller, M. J., \& Kenyon, S. J. 2000, ApJ, 530, 660

\bibitem[Barton Gillespie, Geller, \& Kenyon(2003)]{barton03}
Barton Gillespie, E., Geller, M. J., \& Kenyon, S. J. 2003, ApJ, 582, 668

\bibitem[Barton et al.(2007)]{barton07}
Barton, E. J., Arnold, J. A.,
Zentner, A. R., Bullock, J. S.,
and Wechsler, R. H. 2007, ApJ, 671, 1538

\bibitem[Bergvall, Laurikainen, \& Aalto(2003)]{bergvall03}
Bergvall, N.,
Laurikainen, E. \& Aalto, S. 2003, A\&A 405, 31

\bibitem[Boissier et al.(2007)]{boissier07}
Boissier, S., et al.\ 2007, ApJS, 173, 524


\bibitem[Boselli et al.(2005)]{boselli05}
Boselli, A., et al.\ 2005, ApJ, 623, L13

\bibitem[Boselli et al.(2010)]{boselli10}
Boselli, A., et al.\ 2010, PASP, 122, 261


\bibitem[Bushouse(1987)]{bushouse87}
Bushouse, H. A. 1987, ApJ, 320, 49

\bibitem[Bushouse, Lamb, \& Werner(1988)]{bushouse88}
Bushouse, H. A., Lamb, S. A., \& Werner, M. W. 1988, 
ApJ, 335, 74


\bibitem[Cardelli, Clayton, \& Mathis(1989)]{cardelli89}
Cardelli, J. A., Clayton, G. C., \& Mathis, J. S. 1989, ApJ, 345, 245

\bibitem[Conroy, White, \& Gunn(2010a)]{conroy10a}
Conroy, C., White, M., \& Gunn, J. E. 2010a, ApJ, 708, 58

\bibitem[Conroy, White, \& Gunn(2010b)]{conroy10b}
Conroy, C., White, M., \& Gunn, J. E. 2010b, ApJ, 712, 833

\bibitem[de Vaucouleurs, de Vaucouleurs, \& Corwin(1976)]{deV76}
de Vaucouleurs, G., de Vaucouleurs, A., \& Corwin, H. G. 1976, Second
Reference Catalogue of Bright Galaxies (Austin: University
of Texas Press) (RC2).

\bibitem[de Vaucouleurs(1977)]{deV77}
de Vaucouleurs, G. 1977, in the Proceedings of `The Evolution
of Galaxies and Stellar Populations', eds. B. M. Tinsley and R. B.
Larson (New Haven: Yale University Observatory).

                                                                                


\bibitem[Gil de Paz et al.(2005)]{gildepaz05}
Gil de Paz, A., et al.\ 2005, ApJ,
627, L29

\bibitem[Gil de Paz et al.(2007)]{gildepaz07}
Gil de Paz, A., et al.\ 2007, ApJS,
173, 185

\bibitem[G\'omez et al.(2003)]{gomez03}
G\'omez, P. L., et al.\ 2003, ApJ, 584, 210

\bibitem[Hancock et al.(2007)]{hancock07}
Hancock, M., Smith, B. J., Struck, C., Giroux, M. L.,
Appleton, P. N., Charmandaris, V., \& Reach, W. T. 2007, AJ, 
133, 791

\bibitem[Hancock et al.(2009)]{hancock09}
Hancock, M., Smith, B. J., Struck, C.,
Giroux, M. L, \& Hurlock, S. 2009, AJ,
137, 4643


\bibitem[Keel(2010)]{keel10}
Keel, W. C. 2010, Proceedings
of the `Galaxy Wars: Stellar 
Populations and Star
Formation in Interacting Galaxies' conference, Astronomical
Society of the Pacific Conference Series, Volume 423, 249



\bibitem[Kennicutt et al.(2003)]{kennicutt03}
Kennicutt, R. C., Jr., et al.\ 2003, PASP, 115, 928

\bibitem[Kennicutt et al.(1987)]{kennicutt87}
Kennicutt, R. C., Jr., Keel, W. C., van der Hulst, J. M.,
Hummel, E., \& Roettiger, K. A. 1987, AJ, 93, 1011
                                                                                
\bibitem[Kroupa(2002)]{kroupa02}
Kroupa, P. 2002, Science, 295, 85

\bibitem[Kr\"uger, Fritze-Alvensleben, \& Loose(1995)]{kruger95}
Kr\"uger, H., Fritz-v. Alvensleben, U., \& Loose, H.-H. 1995,
A\&A, 303, 41

\bibitem[Lambas et al.(2003)]{lambas03}
Lambas, D. G., Tissera, P. B., Alonso, M. S., \& Coldwell, G.
2003, MNRAS, 346, 1189

\bibitem[Larson \& Tinsley(1978)]{larson78}
Larson, R. B. \& Tinsley, B. M. 1978, ApJ, 219, 46

\bibitem[Leitherer et al.(1999)]{leitherer99}
Leitherer, C., et al.\ 1999, ApJS, 123, 3

\bibitem[Li et al.(2008)]{li08}
Li, C., Kauffmann, G., Heckman, T. M., Jing, Y. P.,
\& White, S. D. 2008, MNRAS, 385, 1903



\bibitem[Lotz et al.(2006)]{lotz06}
Lotz, J. M., Madau, P., Giavalisco, M.,
Primack, J., \& Ferguson, H. C.
2006, ApJ, 636, 592

\bibitem[Martin et al.(2005)]{martin05}
Martin, D. C., et al.\ 2005, ApJ, 619, L1



\bibitem[M\~noz-Mateos et al.(2009)]{munoz09}
Mu\~noz-Mateos, J. C., et al.\ 2009, ApJ, 703, 1569


\bibitem[Neff et al.(2005)]{neff05}
Neff, S. G., et al.\ 2005, ApJ, 619, L91

\bibitem[Nikolic, Cullen, \& Alexander(2004)]{nikolic04}
Nikolic, B., Cullen, H., \& Alexander, P. 2004, MNRAS, 355, 874

\bibitem[Overzier et al.(2008)]{overzier08}
Overzier, R. A., et al.\ 2008, ApJ, 677, 37

\bibitem[Peterson et al.(2009)]{peterson09}
Peterson, B. W., Struck, C., Smith, B. J., \& Hancock, M.
2009, MNRAS, 400, 1208

\bibitem[Petty et al.(2009)]{petty09}
Petty, S. M., de Mello, D. F., Gallagher,
J. S., Gardner, J. P., Lotz, J. M.,
Matt Mountain, C., \& Smith, L. J.
2009,
AJ, 138, 362


\bibitem[Sandage(1961)]{sandage61}
Sandage, A. 1961, The Hubble Atlas of Galaxies (Washington D.C.:
Carnegie Institute of Washington).

\bibitem[Schlegel, Finkbeiner, \& Davis(1998)]{schlegel98}
Schlegel, D. J., Finkbeiner, D. P., \& Davis, M. 1998, ApJ, 500, 525

\bibitem[Schombert, Wallin, \& Struck-Marcell(1990)]
{schombert90}
Schombert, J. M., Wallin, J. F., \& Struck-Marcell, C. 1990,
AJ, 99, 497

\bibitem[Schweizer(1978)]{schweizer78}
Schweizer, F. 1978, in 
Structure and Properties of Nearby Galaxies,
ed. E. M. Berkhuijsen \& R. Wielebinski (Dordrecht: Reidel), 279

\bibitem[Smith et al.(2008)]{smith08}
Smith, B. J., et al.\ 2008, AJ, 135, 2406

\bibitem[Smith et al.(2010)]{smith10}
Smith, B. J., Giroux, M. L., Struck, C., Hancock, M., \& Hurlock, S.
2010, AJ, 139, 1212; Erratum 2010, AJ, 139, 1212.

\bibitem[Smith et al.(2007)]{smith07}
Smith, B. J., Struck, C., Hancock, M., Appleton, P. N.,
Charmandaris, V., \& Reach, W. T. 2007, AJ, 133, 791

\bibitem[Smith et al.(2005)]{smith05}
Smith, B. J., Struck, C., Appleton, P. N., Charmandaris, V., 
Reach, W., \& Eitter, J. J. 2005, AJ, 130, 2117

\bibitem[Sol Alonso et al.(2006)]{alonso06}
Sol Alonso, M., Lambas, D. G., Tissera, P., \& Coldwell, G. 2006, 
MNRAS, 367, 1029

\bibitem[Struck-Marcell \& Tinsley(1978)]{struckmarcell78}
Struck-Marcell, C. \& Tinsley, B. M. 1978, ApJ, 221, 562

\bibitem[Struck \& Smith(2003)]{struck03}
Struck, C. \& Smith, B. J. 2003, ApJ, 589, 157
                                                                          



                                                                                
\bibitem[V\'{a}zquez \& Leitherer(2005)]{vaz05}
V\'{a}zquez, G. A. \& Leitherer, C. 2005, ApJ, 621, 695



\bibitem[Thilker et al.(2005a)]{thilker05a}
Thilker, D. A., et al.\ 2005a, ApJ, 619, L67

\bibitem[Thilker et al.(2005b)]{thilker05b}
Thilker, D. A., et al.\ 2005b, ApJ, 619, L79


                                                                                
\bibitem[West et al.(2009)]{west09}
West, A. W., Garcia-Appadoo, D. A., Dalcanton, J. J., Disney, M. J.,
Rockosi, C. M., \& Ivezi\'c, Z. 2009, AJ, 138, 796

\bibitem[Witt \& Gordon(2000)]{witt00}
Witt, A. N. \& Gordon, K. D. 2000, ApJ, 463, 681


\end{thebibliography}
\end{document}